\documentclass[12pt]{article}
\usepackage{epsf}
\usepackage[usenames]{color}
\usepackage{hyperref,epsfig,amsmath,times,amssymb,euscript,graphicx,cite,amssymb,empheq}
\usepackage{wrapfig}
\usepackage{pstricks,fancybox,hyperref}
\usepackage{floatflt}
\fontfamily{pnc}\selectfont

\newcommand{\EQ}[1]{\begin{equation} #1 \end{equation}}

\definecolor{myorange}{rgb}{1,0.5,0}
\begin{document}
\setlength{\parskip}{2ex} \setlength{\parindent}{0em}
\setlength{\baselineskip}{3ex}
\newcommand{\onefigure}[2]{\begin{figure}[htbp]
         \caption{\small #2\label{#1}(#1)}
         \end{figure}}
\newcommand{\onefigurenocap}[1]{\begin{figure}[h]
         \begin{center}\leavevmode\epsfbox{#1.eps}\end{center}
         \end{figure}}
\renewcommand{\onefigure}[2]{\begin{figure}[htbp]
         \begin{center}\leavevmode\epsfbox{#1.eps}\end{center}
         \caption{\small #2\label{#1}}
         \end{figure}}
\newcommand{\comment}[1]{}
\newcommand{\myref}[1]{(\ref{#1})}
\newcommand{\secref}[1]{sec.~\protect\ref{#1}}
\newcommand{\figref}[1]{Fig.~\protect\ref{#1}}
\def\sl2z{SL(2,\Z)}
\newcommand{\be}{\begin{equation}}
\newcommand{\ee}{\end{equation}}
\newcommand{\bea}{\begin{eqnarray}}
\newcommand{\eea}{\end{eqnarray}}
\newcommand{\nn}{\nonumber}
\newcommand{\unit}{1\!\!1}
\newcommand{\mt}{\widehat{t}}
\newcommand{\R}{\bf R}
\newcommand{\X}{{\bf X}}
\newcommand{\T}{{\bf T}}
\newcommand{\PP}{\bf P}
\newcommand{\CC}{\bf C}
\newcommand {\us}[1]{\underline{s_{#1}}}
\newcommand {\uk}{\underline{k}}
\newcommand{\me}{\mathellipsis}
\newdimen\tableauside\tableauside=1.0ex
\newdimen\tableaurule\tableaurule=0.4pt
\newdimen\tableaustep
\def\phantomhrule#1{\hbox{\vbox to0pt{\hrule height\tableaurule width#1\vss}}}
\def\phantomvrule#1{\vbox{\hbox to0pt{\vrule width\tableaurule height#1\hss}}}
\def\sqr{\vbox{%
\phantomhrule\tableaustep
\hbox{\phantomvrule\tableaustep\kern\tableaustep\phantomvrule\tableaustep}%
\hbox{\vbox{\phantomhrule\tableauside}\kern-\tableaurule}}}
\def\squares#1{\hbox{\count0=#1\noindent\loop\sqr
\advance\count0 by-1 \ifnum\count0>0\repeat}}
\def\tableau#1{\vcenter{\offinterlineskip
\tableaustep=\tableauside\advance\tableaustep by-\tableaurule
\kern\normallineskip\hbox
    {\kern\normallineskip\vbox
      {\gettableau#1 0 }%
     \kern\normallineskip\kern\tableaurule}%
  \kern\normallineskip\kern\tableaurule}}
\def\gettableau#1 {\ifnum#1=0\let\next=\null\else
  \squares{#1}\let\next=\gettableau\fi\next}

\tableauside=1.0ex \tableaurule=0.4pt

\bibliographystyle{utphys}
\setcounter{page}{1} \pagestyle{plain}
\numberwithin{equation}{section}

\begin{titlepage}
\begin{center}
 \hfill\\ \vskip 1cm {\sc \large Refined Topological vertex, Cylindric Partitions and $U(1)$ Adjoint Theory} \vskip 0.5cm
{\sc Amer\,\,Iqbal$^{1}$ \,\,\,Can Koz\c{c}az$^{2}$\,\,\, Khurram Shabbir$^{3}$}\\
\vskip 0.5cm
$^{1}${ Department of Physics,\\
LUMS School of Science \& Engineering,\\
U Block, D.H.A, Lahore, Pakistan.\\} \vskip 0.5cm
$^{2}${ Department of Physics,\\
University of Washington,\\
Seattle, WA, 98195, U.S.A.\\} \vskip 0.5cm
$^{3}$ {Abdus Salam School
of
Mathematical Sciences,\\
G. C. University,\\
Lahore, Pakistan.}
\end{center}
\vskip 2 cm
\begin{abstract}
We study the partition function of the compactified 5D $U(1)$ gauge theory (in the $\Omega$-background) with a single adjoint
hypermultiplet, calculated using the refined topological vertex. We show that this partition function is an example a periodic Schur process and is a refinement of the generating function of cylindric plane partitions. The size of the cylinder is  given by the mass of adjoint hypermultiplet and the parameters of the $\Omega$-background. We also show that this partition function can be written as a trace of operators which are generalizations of vertex operators studied by Carlsson and Okounkov. In the last part of the paper we describe a way to obtain $(q,t)$ identities using the refined topological vertex.
\end{abstract}
\end{titlepage}
\section{ Introduction}
The topological vertex formalism \cite{Iqbal:2002we,AKMV} has not only been able to completely solve the problem of determining the
Gromov-Witten/Gopakumar-Vafa/Donaldson-Thomas invariants of the toric Calabi-Yau threefolds but has also provided insights into their
combinatorial aspects. In this paper we continue the study of the combinatorial aspects of the Nekrasov's extension of the topological
string partition functions (which are same as the partition functions of the 5D compactified gauge theory in the $\Omega$-background)
for certain toric Calabi-Yau threefolds.  Our main example will be a rather interesting Calabi-Yau threefold $X_{H}$ which gives rise,
via geometric engineering, to $U(1)$ gauge theory with one hypermultiplet in the adjoint representation.
We  provide a combinatorial interpretation of the refined partition function of $X_{H}$ in terms of plane partitions living on a cylinder. These cylindric partitions were studied in \cite{borodin} and are closely related with periodic Schur process. We will see that this cylinder naturally appears in the toric description of $X_{H}$ and the size of the cylinder is determined by the mass of the adjoint hypermultiplet and the parameters $(\epsilon_{1},\epsilon_{2})$ of the $\Omega$-background. We only consider the $U(1)$ theory in this paper, however, the relation with periodic Schur process and cylindric partitions extends to the $U(N)$ theory with adjoint hypermultiplet as well \cite{wip}.

The partition function of the 4D gauge theory was recetly interpreted in terms of matrix elements of a vertex operator corresponding to certain K-theory classes on product of Hilbert schemes of $\mathbb{C}^{2}$ \cite{carlsson}. In this paper we make a similar attempt in trying to understand the compactfied 5D gauge theory partition function in terms of certain $(q,t)$ vertex operators which are a natural generalization of the vertex operators discussed in \cite{carlsson}. The relation with cylindric partitions implies that the matrix elements of these vertex operators are given by number of cylindric partitions of a certain type.

In the last section of the paper we derive a set of $(q,t)$ identities associated
with certain toric CY3-folds. These identities encode the fiber-base duality
of the ${\cal N}=2$ gauge theories \cite{Katz:1997eq}.

 The paper is organized as follows. In section 2 we give a detailed account of the refined vertex using the transfer matrix approach and give the refined crystal picture for the partition function of various CY3-folds. In section 3 we consider the partition function of $U(1)$ adjoint theory in detail and relate it to the combinatorics of cylindric plane partitions. We provide also provide an introduction to the basics of cylindric partitions.  In section 4 we discuss the $(q,t)$ generalization of the vertex operators of \cite{carlsson}. In section 5 we use the choice of the preferred direction needed for the refined vertex calculation to give a set of $(q,t)$ identities associated with certain simple toric CY3-folds.

\section{ 3D Partitions, Refined Vertex and Crystals}

In this section, first we are going to review some background material including the definitions of 2D and 3D partitions, the partition function of a plane partition with multiple variables and the so-called transfer matrix approach to compute those partition functions. We should warn the reader that our presentation is going to be far from the most general form, but rather include only the special cases we need. Later, we are going to focus on the particular parametrization of the refined topological vertex and work out the crystal model for the closed refined topological vertex.

\subsection{ Transfer matrix approach}

 \par{A 2D partition $\nu$ consists of non-negative integers with decreasing order
 $\nu=\{\nu_{1}\geq\nu_{2}\geq\mathellipsis,|\nu_{i}\geq0\}$. The pictorial representation obtained by
 placing $\nu_{i}$ boxes next to each other relates them to the Young diagrams. If we have another
 2D partition $\lambda$ in addition to $\nu$ such that $\lambda_{i}\geq\nu_{i}$ for all $i$ we say
 $\lambda$ includes $\nu$ and denote it by $\nu\subseteq\lambda$. This condition implies that any
 box $(i,j)$ belonging to $\nu$ is also an element of $\lambda$. For two such partitions we can
 construct the skew partition $\lambda/\nu$ by removing all boxes that are elements of $\nu$ from
 $\lambda$, \textit{i.e.}, $\lambda/\nu=\{(i,j)\in\lambda|(i,j)\notin\nu\}$. It is obvious from this definition
 that a skew partition might not be a partition. However, if $\lambda$ is chosen to be
 $\lambda=\{\lambda_{i}=L \,|\, i=1,\mathellipsis,M\}$ such that $L$ and $M$ are large enough to
 include $\nu$ in $\lambda$, the skew partition $\lambda/\nu$ is always a 2D partition.}

A plane partition $\pi$ is defined by an array of non-negative integers satisfying
 \bea \label{planepartition}
 \pi_{i,j}\geq\pi_{i+r,j+s}\,,\,\,\,\,r,s\geq 0.
\eea
Plane partitions have also a 3-dimensional pictorial representation: if we divide the base
$xy$-plane into unit squares and denote them by $(i,j)$, we can place $\pi_{i,j}$ boxes over each
square $(i,j)$. In this sense, plane partitions are considered a generalization of Young diagrams.
The total number of boxes of a plane partition $\pi$ is given by
\bea\nn
 |\pi|=\sum_{i,j}\pi_{i,j}.
 \eea 
A skew 3D partition of shape $\lambda/\nu$ is an array of
non-negative integers $\{\pi_{i,j}\,\,|\,\,(i,j)\in \lambda/\nu\}$ satisfying the same condition as in Eq. \eqref{planepartition}. 
\vskip5cm
\begin{wrapfigure}[10]{l}[0in]{2.0in}
\centering \epsfxsize=2.0in \epsffile{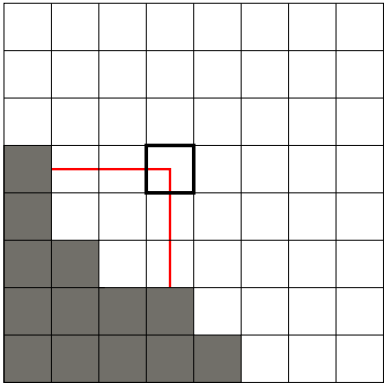} \caption{\small The hook length of a box in $\nu^{c}$.} \label{hook}
\end{wrapfigure}
 The partition function corresponding to a skew plane partition of shape $\nu^{c}$ (the complement of $\nu$) is given by
\bea
Z_{\nu}(q):&=&\sum_{\pi(\nu^{c})}q^{|\pi|}\\ \nonumber
&=&\prod_{(i,j)\in\nu^{c}}\frac{1}{1-q^{{\hat h}(i,j)}},
\eea
where ${\hat h}=j-\nu_{i}+i-\nu_{j}^{t}-1$ is the hook length of the box $(i,j)\in\nu^{c}$ as shown in \figref{hook}.
\\
\\

If $\nu$ is the empty partition $\emptyset$, the partition function becomes the MacMahon function,
\bea
M(q):=Z_{\emptyset}(q)=\prod_{k=1}^{\infty}\frac{1}{(1-q^{k})^{k}}.
\eea
If we normalize the partition function $Z_{\nu}(q)$ by the MacMahon function $M(q)$ then we obtain ${\widetilde Z}_{\nu}(q)$ given by
\bea
{\widetilde Z}_{\nu}(q):=\frac{Z_{\nu}(q)}{M(q)}=\prod_{(i,j)\in\nu}\frac{1}{1-q^{h(i,j)}},
\eea
where $h(i,j)=\nu_{i}-j+\nu_{j}^{t}-i+1$ is the hook length of a box in $\nu$.
This partition function as well as more general ones which we will define shortly can be computed using transfer matrix formalism. We can consider a plane partition function as a sequence of 2D partitions, $\{\eta(a)|a\in \mathbb{Z}\}$, along the slices whose projection to the base is given by $y-x=a$.

\begin{floatingfigure}[l]{2.4in}
\centering \epsfxsize=2.2in \epsffile{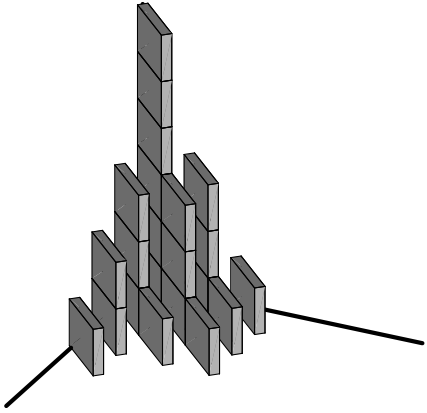} \caption{\small The diagonal slicing of the plane partition.} \label{slice}
\end{floatingfigure}
This sequence is obtained by slicing the plane partition by diagonal planes as shown in \figref{slice}. The definition of a plane partition puts strong conditions among the 2D partitions in the sequence. Before we spell out these conditions we need to define \textit{interlacing}: we say a 2D partition $\mu$ interlaces another 2D partition $\nu$, written as $\mu\succ\nu$, if
\bea
\mu_{1}\geq\nu_{1}\geq\mu_{2}\geq\nu_{2}\geq\mathellipsis\,.
\eea
Note that interlacing is a stronger condition than including. The diagonal slices $\eta(a)$ obtained from a plane partition satisfy
\bea\label{interlacing}
\eta(a+1)&\succ&\eta(a),\,\,\,\,\,\,\,\,\,\,\,\,\,\, a<0, \\ \nonumber
\eta(a)&\succ&\eta(a+1), \,\,\,\,\, a\geq0.
\eea

Now we are ready to define, following \cite{Okounkov}, a more generalized partition function of a plane partition: we can weigh different slices of the 3D partition with different variables, hence the partition function becomes
\bea
Z_{\nu}(\{q_{a}\}):=\sum_{\{\eta(a)\}}\prod_{a\in\mathbb{Z}}q_{a}^{|\eta(a)|}=\sum_{\pi}\prod_{a\in\mathbb{Z}}q_{a}^{|\pi_{a}|}
\eea

where $|\pi_{a}|=\sum_{i}\pi_{i,a+i}$. It is obvious from this definition that we will obtain the partition function we previously defined if we set all variables equal to each other, $\{q_{a}=q\}$.
\par{The transfer matrix approach is based on associating a fermionic Fock space to 2D partitions. We will introduce creation/annihilation as well as the so-called vertex operators acting on this space of states equipped with a natural inner product. }
\par{The Fock space $\mathcal{F}$ is a semi-infinite product of another vector space $V$ spanned by vectors $\uk$, where $k\in\mathbb{Z}+1/2$. An element $v_{S}$ of the Fock space ${\mathcal F}=\bigwedge^{\frac{\infty}{2}}V$ is given by}
\bea
v_{S}=\us1\wedge\us2\wedge\us3\wedge\mathellipsis\, ,
\eea
where $S=s_{1}>s_{2}>\mathellipsis$ is a subset of $\mathbb{Z}+1/2$, such that $S\,\backslash\, (\mathbb{Z}_{\leq0}-1/2)$ and $(Z_{\leq0}-1/2)\,\backslash\, S$ are both finite. Over this space there is a natural inner product with respect to which the basis defined by $\{v_{S}\}$ is orthonormal.

\par{The generators $\psi_{k}$ and $\psi_{k}^{*}$ of Clifford algebra satisfy the following anti-commutation relations:}
\bea
\{\psi_{k},\psi_{k'}\}=0,\,\,\,\, \{\psi_{k}^{*},\psi_{k'}^{*}\}=0, \,\,\,\, \{\psi_{k},\psi_{k'}^{*}\}=\delta_{kk'},
\eea
where $k,k'\in\mathbb{Z}+1/2$. Later we will need the explicit action of the Clifford algebra generators on the basis vectors
\bea
\psi_{k}\left ( \us1\wedge\us2\wedge\us3\wedge\mathellipsis\right)&=&\uk\wedge\us1\wedge\us2\wedge\us3\wedge\mathellipsis \\
\psi_{k}^{*}\left ( \us1\wedge\us2\wedge\mathellipsis\wedge\us l\wedge \uk\wedge\us {l+1}\wedge\mathellipsis\right)&=&(-1)^{l}\us1\wedge\us2\wedge\mathellipsis\wedge\us l\wedge\us {l+1}\wedge\mathellipsis \\
\psi_{k}^{*}\left(\us1\wedge\us2\wedge\us3\mathellipsis \right)&=&0,\,\,\,\, \hbox{for}\, k\in\mathbb{Z}\backslash S.
\eea
The vectors in the Fock space $\mathcal{F}$ can be parameterized by partitions:
\bea
v_{\lambda}^{(0)}=\underline{\lambda_{1}-1/2}\wedge\underline{\lambda_{2}-3/2}\wedge\mathellipsis.
\eea
where we ignore an irrelevant shift $m$ of the vacuum energy from the definition in \cite{Okounkov}. In our notation, $v_{0}^{(0)}$ denotes the vacuum state corresponding to empty partition $\emptyset$.

\par{At this point, we have constructed a fermionic Fock space with Clifford algebra acting on it and established one-to-one correspondence with the 2D partition. What is needed to continue is an operator which can create states corresponding to 2D partitions which interlace a given 2D partition after  acting on a given  state. To construct this operator first we need to define $\alpha_{n}$ }
\bea
\alpha_{n}=\sum_{k\in\mathbb{Z}+1/2}\psi_{k+n}\psi_{k}^{*},\,\,\,\,\,\,\,\, n=\pm1,\pm2,\mathellipsis\,.
\eea
These operators satisfy the following commutation relationships:
\bea
[\alpha_{n},\alpha_{m}]=-n\delta_{n,-m},\,\,\,\,\,[\alpha_{n},\psi_{k}]=\psi_{k+n},\,\,\,\,\,[\alpha_{n},\psi_{k}^{*}]=-\psi_{k-n}^{*}.
\eea
The vertex operators are obtained from these operators $\alpha_{n}$'s by exponentiation
\bea
\Gamma_{+}(x)&=&\exp\left(\sum_{n\geq1}\frac{x^n}{n}\alpha_{n}\right),\\
\Gamma_{-}(x)&=&\exp\left(\sum_{n\geq 1}\frac{x^n}{n}\alpha_{-n}\right).
\eea
$\Gamma_{+}(x)$ and $\Gamma_{-}(x)$ are conjugates of each other with respect to the inner product on $\mathcal{F}$:
\bea\label{conjugate}
\left(\Gamma_{-}(x)v,w\right)=\left(v,\Gamma_{+}(x)w\right).
\eea
The action of $\Gamma_{-}(x)$ on the vacuum state is particularly important, $\Gamma_{-}(x)v_{0}^{(0)}=v_{0}^{(0)}$. It is easy to verify that they satisfy the following commutation relation which we are going to use extensively (in addition to its action on the vacuum state as well as the conjugacy property):
\bea\label{commutation}
\Gamma_{+}(x)\Gamma_{-}(y)=(1-xy)\Gamma_{-}(y)\Gamma_{+}(x).
\eea
Let us discuss this formal construction a little bit more explicitly. For the simplest cases, one can  easily  convince oneself of the above relation by expanding the exponential in $\Gamma_{+}(x)$ as a power series and acting with the individual terms in the expansion on a given state. One creates a generating function for all partitions that interlace the one acted on. This is summarized as
\bea
\prod_{i}\Gamma_{+}(x_{i})v_{\mu}^{(0)}=\sum_{\lambda\supset\mu}s_{\lambda/\mu}(x)v_{\lambda}^{(0)}.
\eea
More specifically
\bea
\Gamma_{+}(1)v_{\mu}^{(0)}=\sum_{\lambda\succ\mu}v_{\lambda}^{(0)},
\eea
since $s_{\lambda/\mu}(1)=1$ if $\lambda\succ\mu$, and vanishes otherwise.

\par{The generalized partition function can be written in terms of the vertex operators. To sum over all possible plane partitions, one way is to start at $a=\infty$ with vacuum and apply $\Gamma_{+}(x)$. We end up with all possible partitions as a generating function on the next slice that interlace vacuum. Then we successively apply $\Gamma_{+}(x)$ until we hit the main diagonal slice $a=0$. This way, we create partitions which interlace the partitions on the previous slice. After the main diagonal we start applying $\Gamma_{-}(y)$'s successively, and  we create partitions that are interlaced by the previous slices, until we reach at $a=-\infty$. The transfer matrix formalism can be used in a more general situation when we asymptotically have non-trivial states at $a=\pm\infty$. }

\begin{wrapfigure}{r}{3.15in}
\centering \epsfxsize=2.8in \epsffile{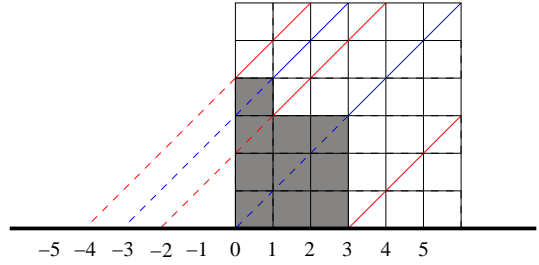} \caption{\small Three inner corners $\{v_{1},v_{2},v_{3}\}=\{-4,-2,3\}$, two outer corners $\{u_{1},u_{2}\}=\{-3,0\}$.} \label{corners}
\end{wrapfigure}

The partition function of a skew 3D partition depends on the 2D partition $\nu$ on the base. We divide the corners of the corresponding 2D partition into \textit{inner} and \textit{outer} corners. We parameterize the inner and outer corners by their coordinates $v_{i}$  and $u_{i}$, respectively, of their projection onto the real line as shown in \figref{corners}. It is convenient to introduce another set of parameters $\{x_{m}^{\pm}|m\in\mathbb{Z}+1/2\}$ and identify them with $q_{a}$'s in the following shape dependent way \cite{Okounkov}:
\vspace{0.4in}
\bea \label{trans}
\frac{x^{+}_{m+1}}{x^{+}_{m}}&=&q_{m+\frac{1}{2}}\,,\,\,\,\,m>v_{M}\,\,\mbox{or}\,\,\,u_{i}-1>m>v_{i}\,,\\ \nn
x^{+}_{u_{i}-\frac{1}{2}}x^{-}_{u_{i}+\frac{1}{2}}&=&
q_{u_{i}}^{-1}\,,\\ \nn
x^{-}_{v_{i}-\frac{1}{2}}x^{+}_{v_{i}+\frac{1}{2}}&=&q_{v_{i}}\,,\\\nn
\frac{x^{-}_{m}}{x^{-}_{m+1}}&=&q_{m+\frac{1}{2}}\,,\,\,m<v_{1}\,\,\mbox{or}\,\,\,v_{i+1}-1>m>u_{i}\,,
\eea
where $M$ is the number of outer corners. In terms of these new variables, the generalized partition function reads
\begin{align}
Z_{\nu}(\{x^{\pm}_{m}\})=&\left( \prod_{u_{M}>m>v_{M}}\Gamma_{-}(x^{+}_{m}) \mathellipsis \prod_{u_{i}<m<v_{i+1}}\Gamma_{+}(x^{-}_{m})\prod_{v_{i}<m<u_{i}}\Gamma_{-}(x^{+}_{m})\mathellipsis \right. \\ \nonumber &\left. \prod_{v_{1}>m>u_{0}}\Gamma_{+}(x^{-}_{m})v_{0}^{(0)},v_{0}^{(0)} \right)=\left( \prod_{u_{0}<m<u_{N}}\Gamma_{-\varepsilon(m)}(x^{\varepsilon(m)}_{m})v_{0}^{(0)},v_{0}^{(0)} \right),
\end{align}
where $m$ runs over $\mathbb{Z}+1/2$. In the last equation, we have introduced some new notation; $\varepsilon(m)=+$, if $v_{i}<m<u_{i}$ for $1\leq i\leq M$ and, $\varepsilon(m)=-$, if $u_{i}<m<v_{i+1}$ for $0\leq i\leq M-1$.
This partition function turns out to have a nice compact form \cite{Okounkov}
\begin{equation}
 Z_{\nu}(\{x^{\pm}_{m}\})=\prod_{\substack{m_{1}<m_{2}\\m_{1}\in D^{-},m_{2}\in D^{+}}}\left(1-x^{-}_{m_{1}}x^{+}_{m_{2}}\right)^{-1},
 \end{equation}
with $D^{\pm}=\{m\,|\,\varepsilon(m)=\pm\}$.

\subsection{ Refined topological vertex}
\par{The transfer matrix formalism reviewed in the last section is capable of computing a partition function with infinitely many parameters, one for each diagonal slice. However, the main motivation to construct a more refined topological vertex comes from the microscopic derivation of the Seiberg-Witten solution \cite{Nekrasov:2002qd}, and it has only two distinct parameters, $q$ and $t$. We will review the correct choice of assigning equivariant parameters to the diagonal slices, \textit{i.e.,} the map $\{q_{a}|a\in\mathbb{Z}\}\mapsto\{q,t\}$. We will refer the interested reader to the original reference \cite{IKV} for the details of the physical motivation as well as the derivation of the refined topological vertex. Here, we only give the map.}
\begin{wrapfigure}{r}{2.3in}
\centering \epsfxsize=2.3in \epsffile{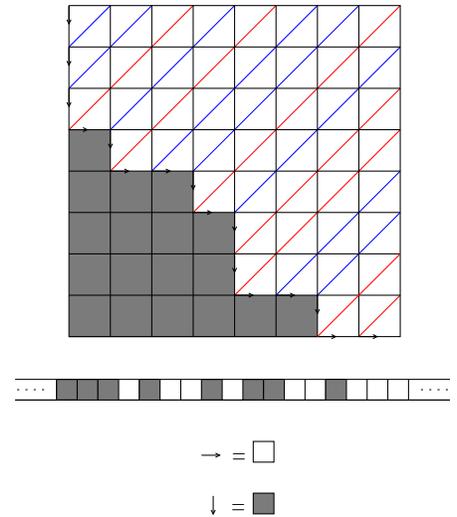} \caption{\small The parameters $q$ and $t$ are assigned based on the partition $\nu=(5,4,4,3,1,1)$.} \label{trace}
\end{wrapfigure}\\
Imagine we are computing the partition function $Z_{\nu}$ for a partition $\nu$ shown in \figref{trace}. According the transfer matrix method we have a series of diagonal slices presented by the red and blue lines in the figure. As we mentioned before we start at $a=\infty$ and apply $\Gamma_{\pm}$ repeatedly, following the arrows. For each partition we can construct a ``barcode'' by assigning a black or white box depending on whether we are going horizontally or vertically, respectively, while we are tracing the profile of $\nu$. Note that if we count the number of black boxes to the left of the $i^{th}$ white box, we get $\nu_{i}$. For example, there are 5 black boxes to the left of the first white box. Similarly, if we count the number of white boxes to the right of the $j^{th}$ black box, we obtain $\nu_{j}^{t}$. It turns out that this barcode is in one-to-one correspondence with 2D partitions.  The $(q,t)$-assignment to each slice is essentially determined by this barcode, whenever we go vertically we count that slice with $q$, otherwise, along each horizontal pass, with $t$. For an arbitrary 2D partition $\nu$ the map has the following form:
\bea\label{map}
\{x_{m}^{+}\,|\,m\in D^{+}\}&=&\{t^{i}q^{-\nu_{i}}\,|\,i=1,2,3,\mathellipsis\},\\
\{x_{m}^{-}\,|\,m\in D^{-}\}&=&\{q^{j-1}t^{-\nu_{j}^{t}}\,|\,j=1,2,3,\mathellipsis\}.
\eea
In general, at the asymptotes $a=\pm\infty$, we may not have empty partitions but instead two different partitions, say $\lambda$ and $\mu$. The interlacing condition Eq. (\ref{interlacing}) requires us to excise the whole region behind those two partitions. In this sense, we are computing the transition from a 2D partition $\lambda$ to another one $\mu$. This is the generating function of all possible 3D partitions we can create by putting boxes in the empty region left by excising the three asymptotes \cite{ORV},
\bea
Z_{\lambda\,\mu\,\nu}(t,q):=
\left( \prod_{u_{0}<m<u_{N}}\Gamma_{-\varepsilon(m)}(x^{\varepsilon(m)}_{m})v_{\lambda}^{(0)},v_{\mu}^{(0)} \right)\,.\eea
The refined topological vertex is then defined to be
\bea
C_{\lambda\,\mu\,\nu}(t,q)&=&\frac{Z_{\lambda\,\mu\,\nu}(t,q)}{Z_{\emptyset\,\emptyset\,\emptyset}(t,q)}\\\nn
&=&\Big(\frac{q}{t}\Big)^{\frac{||\mu||^2+||\nu||^2}{2}}\,t^{\frac{\kappa(\mu)}{2}}\,
P_{\nu^{t}}(t^{-\rho};q,t)\,
\sum_{\eta}\Big(\frac{q}{t}\Big)^{\frac{|\eta|+|\lambda|-|\mu|}{2}}s_{\lambda^{t}/\eta}(t^{-\rho}q^{-\nu})
s_{\mu/\eta}(t^{-\nu^{t}}q^{-\rho}),
\label{RV}
\eea
where
\bea\nn
P_{\nu^t}(t^{-\rho};q,t)&=&t^{\frac{||\nu||^2}{2}}\,\widetilde{Z}_{\nu}(t,q)\\\nn
&=&t^{\frac{||\nu||^2}{2}}\,\prod_{(i,j)\in\nu}\Big(1-t^{a(i,j)+1}\,q^{\ell(i,j)}\Big)^{-1}\,,\,\,a(i,j)=\nu^{t}_{j}-i\,,\,\,\ell(i,j)=\nu_{i}-j\,.
\eea
\subsection{ Crystal models and $(t,q)$ parameters}

In this section, we will discuss crystal models for $X_{0}:={\cal O}(-1)\oplus {\cal O}(-1)\mapsto \mathbb{P}^{1}$ and $X_{1}:={\cal O}(0)\oplus {\cal O}(-2)\mapsto \mathbb{P}^{1}$. We will see that both these models have exactly the same combinatorial description with the only difference being the expansion parameters. A better understanding of these combinatorial models and the expansion parameters $(q,t)$ will help us later understand the combinatorial model for the compactified resolved conifold which gives rise to $U(1)$ gauge theory with adjoint matter.

Recall that the refined partition function of $X_{0}$ and $X_{1}$ is given by \cite{IKV} (this can be calculated using the topological vertex formalism)\footnote{The refined topological string partition function in terms of Gopakumar-Vafa invariants can be written as
\bea\nn
Z_{X}(\omega,t,q):=\prod_{C\in H_{2}(X,Z)}\prod_{j_{L},j_{R}}\prod_{k_{L}=-j_{L}}^{j_{L}}\prod_{k_{R}=-j_{R}}^{j_{R}}\prod_{m_{1},m_{2}=1}^{\infty}
(1-t^{k_{L}+k_{R}+m_{1}-\frac{1}{2}}\,q^{k_{L}-k_{R}+m_{2}-\frac{1}{2}}\,e^{-\omega\cdot C})^{(-1)^{2(j_{L}+j_{R})}N_{C}^{j_{L},j_{R}}(X)}\,.
\eea
For the case of $X_{0}$ and $X_{1}$ this simplifies to
\bea
Z_{X_{k}}(Q,t,q)=\prod_{j_{L},j_{R}}\prod_{k_{L}=-j_{L}}^{j_{L}}\prod_{k_{R}=-j_{R}}^{j_{R}}\prod_{m_{1},m_{2}=1}^{\infty}
(1-t^{k_{L}+k_{R}+m_{1}-\frac{1}{2}}\,q^{k_{L}-k_{R}+m_{2}-\frac{1}{2}}\,Q)^{(-1)^{2(j_{L}+j_{R})}N_{C}^{j_{L},j_{R}}(X_k)}\,.
\eea
For $X_{0}$ the moduli space of $\mathbb{P}^{1}$ is just a point therefore $N^{j_L,j_R}_{C}=\delta_{j_{L},0}\delta_{j_{R},0}$. For $X_{1}$ the moduli space of $\mathbb{P}^{1}$ is $\mathbb{C}$. If the moduli space had been $\mathbb{P}^{1}$ this would have given the spin content $(j_{L},j_{R})=(0,\frac{1}{2})\Rightarrow (k_{L},k_{R})=(0,\{-\frac{1}{2},+\frac{1}{2}\})$. Since the moduli space is $\mathbb{C}$ we can think of it as half-$\mathbb{P}^{1}$ giving the spin content $(j_{L},j_{R})=(0,\frac{1}{2})\Rightarrow (k_{L},k_{R})=(0,\{+\frac{1}{2}\})$. Thus we get
\bea
Z_{X_{0}}(Q,t,q)=\prod_{i,j=1}^{\infty}\Big(1-Q\,t^{i-\frac{1}{2}}\,q^{j-\frac{1}{2}}\Big)\,,\,\,\,
Z_{X_{1}}(Q,t,q)=\prod_{i,j=1}^{\infty}\Big(1-Q\,t^{i}\,q^{j-1}\Big)^{-1}\,.
\eea
Of course, we could choose the spin content for half-$\mathbb{P}^{1}$ to be $(j_{L},j_{R})=(0,\frac{1}{2})\Rightarrow (k_{L},k_{R})=(0,\{-\frac{1}{2}\})$. In this case we get
\bea
Z_{X_{1}}(Q,t,q)=\prod_{i,j=1}^{\infty}\Big(1-Q\,t^{i-1}\,q^{j}\Big)^{-1}\,.
\eea
In terms of combinatorics of 3D partitions the two choices for the spin content correspond to the choice of counting the partition $\eta(0)$ (the 2D partition on the main diagonal) with $t$ or $q$.
}
\bea
Z_{X_{0}}(Q,t,q)=\prod_{i,j=1}^{\infty}\Big(1-Q\,t^{i-\frac{1}{2}}\,q^{j-\frac{1}{2}}\Big)\,,\,\,\,
Z_{X_{1}}(Q,t,q)=\prod_{i,j=1}^{\infty}\Big(1-Q\,t^{i}\,q^{j-1}\Big)^{-1}\,.
\eea
Where $T=-\mbox{log}(Q)$ is the K\"ahler parameter associated with the $\mathbb{P}^{1}$ in the geometry.
The combinatorial interpretation of $Z_{X_{1}}(Q,q,q)$ in terms of 3D partitions is well known \cite{Maeda:2004iq}. A similar combinatorial interpretation for $Z_{X_{1}}(Q,t,q)$ can be found: Given a 3D partition $\pi$ and its diagonal slices $\{\eta(a)\,,a\in \mathbb{Z}\}$
\bea
\sum_{\pi,\eta(0)=\lambda}\,q^{\sum_{a>0}|\eta(a)|}\,t^{\sum_{a\leq 0}|\eta(a)|}&=&t^{\sum_{i}i\lambda_{i}}\,q^{\sum_{i}(i-1)\lambda_{i}} \prod_{s\in \lambda}\Big(1-q^{h(s)}\Big)^{-1}\,\Big(1-t^{h(s)}\Big)^{-1}\\\nn
&=&s_{\lambda}(t,t^{2},t^{3},\cdots)\,s_{\lambda}(1,q,q^{2},\cdots)\,.
\eea
The prefactor $t^{\sum_{i}i\lambda_{i}}\,q^{\sum_{i}(i-1)\lambda_{i}}$ arises because if $\eta(0)=\lambda$ then the 3D partition with the least number of boxes is such that there are $\sum_{i}i\lambda_{i}$ number of boxes on or to the right of the main diagonal and $\sum_{i}(i-1)\lambda_{i}$ is the number of boxes below the diagonal. Thus
\bea\nn
\sum_{\lambda}Q^{|\lambda|}\sum_{\pi,\eta(0)=\lambda}\,q^{\sum_{a>0}|\eta(a)|}\,t^{\sum_{a\leq 0}|\eta(a)|}&=&\sum_{\lambda}Q^{|\lambda|}s_{\lambda}(t,t^2,\cdots)\,s_{\lambda}(1,q,\cdots)\\\nn
&=&\prod_{i,j=1}^{\infty}\Big(1-Q\,t^{i}\,q^{j-1}\Big)^{-1}=Z_{X_{1}}(Q,t,q)\,.
\eea
The refined partition function $Z_{X_{0}}(Q,t,q)$ has a similar combinatorial description in which instead of counting the slices with parameters $q$ and $t$ we count them with $q$ and $t^{-1}$ and also splitting the slice $\eta(0)$ symmetrically between the two parameters,
\bea\nn
&&\sum_{\lambda}Q^{|\lambda|}\sum_{\pi,\eta(0)=\lambda}\,q^{\frac{|\eta(0)|}{2}+\sum_{a>0}|\eta(a)|}\,(t^{-1})^{\frac{|\eta(0)|}{2}
+\sum_{a< 0}|\eta(a)|}=\sum_{\lambda}Q^{|\lambda|}s_{\lambda}(t^{-\frac{1}{2}},t^{-\frac{3}{2}},\cdots)\,
s_{\lambda}(q^{\frac{1}{2}},q^{\frac{3}{2}},\cdots)\\\nn
&=&\sum_{\lambda}Q^{|\lambda|}(-1)^{|\lambda|}s_{\lambda^t}(t^{\frac{1}{2}},t^{\frac{3}{2}},\cdots)\,
s_{\lambda}(q^{\frac{1}{2}},q^{\frac{3}{2}},\cdots)=\prod_{i,j=1}^{\infty}\Big(1-Q\,t^{i-\frac{1}{2}}\,q^{j-\frac{1}{2}}\Big)=Z_{X_{0}}(Q,t,q)\,.
\eea
Thus we see that both $Z_{X_{0}}(Q,t,q)$ and $Z_{X_{1}}(Q,t,q)$ can be expressed as a sum over 3D partitions as long as correct expansion parameters are chosen.

\begin{wrapfigure}{l}{3in}
\centering \epsfxsize=3in \epsffile{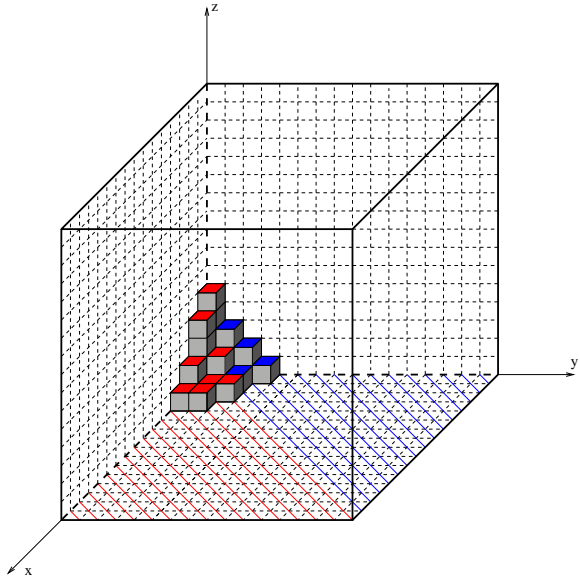} \caption{\small 3D partition and choice of slicing.} \label{3Dcrystal}
\end{wrapfigure}

The above is not the only crystal model for $Z_{X_{0}}(Q,q,q)$. Another model in terms 3D partitions has been discussed \cite{Okuda:2004mb}. The model consists of putting an additional ``wall'' parallel to one of the already existing walls bounding the positive octant ${\mathbb R}^{3+}$, say the one along the $xz$-plane, in the region where we are growing our crystal. The distance of this additional wall to $xz$-plane is related to the K\"{a}hler parameter of the resolved conifold. The region bounded by these walls seems like the toric diagram of the resolved conifold. Later we will consider the double-$\mathbb{P}^{1}$ and the closed topological vertex for the refined case. First, we will allow the size of the preferred direction to be non-compact. This is equivalent to placing another wall, now parallel to the $yz$-plane, and allowing the crystal grow in the $z$-direction without any bound. Again, the location of this second wall is related to the appropriate K\"{a}hler parameter in the toric geometry. Later, for the closed refined topological vertex, we will introduce a projection operator and put a ``ceiling'' in the region where we grow the crystal, that is equivalent putting a last wall parallel to the $xy$-plane.

In \figref{3Dcrystal},  we show an example of a plane partition and show how the slices should be weighted; each blue slice, \textit{i.e.} $a>0$,  gives rise to a factor of $q^{|\eta(a)|}$, whereas each red slice contributes $t^{|\eta(a)|}$ with $a\leq0$. For instance, the 3D partition in the figure counts as $q^{7}t^{17}$.

The refined partition function of $X_{0}$ can also be obtained by putting a wall at a distance of $M$ along the $y$-direction. The partition function reads then
\bea
Z_{\mbox{\tiny crystal}}=\left( \prod_{\infty>m>0}\Gamma_{-}(x^{+}_{m})\prod_{0>m>-M}\Gamma_{+}(x_{m}^{-})v_{0}^{(0)},v_{0}^{(0)}\right).
\eea
We can repeatedly make use of the commutation relation Eq.(\ref{commutation}) of the vertex operators to get
\bea
Z_{\mbox{\tiny crystal}}=\prod_{k_{1}=1}^{\infty}\prod_{k_{2}=1}^{M} \left(1-x_{k_{1}-1/2}^{+}x_{-k_{2}+1/2}^{-} \right)^{-1} \underbrace{\left( \prod_{0>m>-M}\Gamma_{+}(x_{m}^{-})\prod_{\infty>m>0}\Gamma_{-}(x_{m}^{+})v_{0}^{(0)},v_{0}^{(0)}\right )}_{=1}
\eea
and then it is easy to see that the inner product is equal to 1, due to the Eq. (\ref{conjugate}) and the fact that $\Gamma_{-}(x_{m}^{+})$ acts as identity on the vacuum state $v_{0}^{(0)}$. We have already established the map between $\{q,t\}$ and $\{x_{m}^{\pm}\}$ in Eq. (\ref{map}), the partition function from the 3D crystal takes the form
\bea
Z_{\mbox{\tiny crystal}}=\prod_{i=1}^{\infty}\prod_{j=1}^{M}\left(1-t^{i}q^{j-1}\right)^{-1}.
\eea
Since
\bea
Z_{X_{0}}(Q,t,q)=\prod_{i=1}^{\infty}\prod_{j=1}^{\infty}\left(1-Q t^{i-\frac{1}{2}}q^{j-\frac{1}{2}}\right),
\eea
these two partition functions turn out to be related to each other in the same way as in \cite{Sulkowski:2006jp}:
\bea\label{vc}
Z_{\mbox{\tiny crystal}}=M(t,q)Z_{X_{0}}(Q,t,q),
\eea
with the identification $Q\sqrt{\frac{q}{t}}=q^{M}$. $M(t,q)$ is the refined MacMahon function already defined in \cite{IKV}:
\bea
M(t,q)=\prod_{i=1}^{\infty}\prod_{j=1}^{\infty}\left(1-t^{i}q^{j-1}\right)^{-1}.
\eea
\newpage
\subsubsection{Double-$\mathbb{P}^{1}$ and closed refined topological vertex}
In this section, we will first place the second wall in our crystal and then put the ceiling. While introducing the second wall is in the same spirit as placing the first wall, the ceiling will require, as mentioned, the introduction of a projection operator ${\mathcal P}_{N}$.

\begin{floatingfigure}[r]{2.0in}
\centering \epsfxsize=2.0in \epsffile{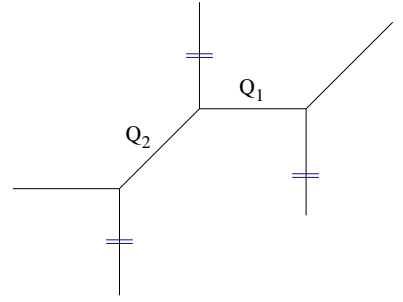} \caption{\small Toric diagram of double-$\mathbb{P}^{1}$.} \label{closed2}
\end{floatingfigure}
In \figref{closed2} we have the toric diagram for the double-$\mathbb{P}^{1}$. The double blue lines show our choice of the preferred direction. The crystal partition function is given by
\bea
Z_{\mbox{\tiny crystal}}=\left( \prod_{L>m>0}\Gamma_{-}(x^{+}_{m})\prod_{0>m>-M}\Gamma_{+}(x_{m}^{-})v_{0}^{(0)},v_{0}^{(0)}\right)
\eea
We repeat the same steps as in the previous example and obtain
\bea
Z_{\mbox{\tiny crystal}}=\prod_{i=1}^{L}\prod_{j=1}^{M}(1-t^{i}q^{j-1})^{-1}.
\eea
The refined vertex computation (see Appendix A) gives
\bea
Z_{\mbox{\tiny double}\,\mathbb{P}^{1}}(Q_{1},Q_{2},t,q)=\prod_{i=1}^{\infty}\prod_{j=1}^{\infty}\frac{(1-Q_{1}t^{i-\frac{1}{2}}q^{j-\frac{1}{2}})(1-Q_{2}t^{i-\frac{1}{2}}
q^{j-\frac{1}{2}})}{(1-Q_{1}Q_{2}t^{i-1}q^{j})}.
\eea
After the following identification
\bea\nn
Q_{1}\sqrt{\frac{q}{t}}=q^{M},\,\,\,
Q_{2}\sqrt{\frac{q}{t}}=t^{L},
\eea
we get
\bea
Z_{\mbox{\tiny crystal}}=M(t,q)\,Z_{\mbox{\tiny double}\,\mathbb{P}^{1}}(Q_{1},Q_{2},t,q).
\eea
Let us now introduce the projection operator we mentioned before.
The goal for such an operator is to eliminate the contributions
coming from the 3D partition which are higher than our ``ceiling''.
The projection operator can be written in terms of the fermionic
operators $\psi_{k}$ and $\psi^{*}_{k}$. Using the fact that
\cite{Okounkov}\bea \psi^{*}_{k}\psi_{k}|\lambda \rangle =\left\{
                                        \begin{array}{ll}
                                          0 & \mbox{if $k=\lambda_{i}-i+\frac{1}{2}$\,,\,\,for some $i=1,2,3,\cdots$} \\
                                          |\lambda\rangle & \mbox{otherwise}
                                        \end{array}
                                      \right.
 \eea it is easy to see that the projection operator is given by
\bea {\mathcal P}_{N}=\prod_{j=N+\frac{1}{2}}^{\infty}\psi^{*}_{j}\psi_{j}\,.
\eea For the purpose of calculating the generating function it is
more useful to write the above projection operator as \bea
{\mathcal P}_{N}=\prod_{j=N+\frac{1}{2}}^{\infty}\psi^{*}_{j}\psi_{j}=\sum_{\eta,\eta_{1}\leq
N}|\eta\rangle \langle \eta|\,. \eea Thus the refined generating
function of the 3D partitions in a $L\times M\times \infty$ box with
the restriction that the 2D partition on the diagonal slice through
the origin is $\eta$. Thus if we want to put a ceiling of height $N$,
we can do this by allowing only those partitions $\eta$ for which
$\eta_{1}\leq N$. Thus the generating function of the 3D partitions
in a $L\times M\times N$ box is \bea Z(N,M,L)=\sum_{\eta,\eta_{1}\leq
N}\langle 0| \prod_{L>m>0}\Gamma_{-}(x^{+}_{m})|\eta \rangle\langle
\eta|\prod_{0>m>-M}\Gamma_{+}(x_{m}^{-})|0\rangle.\,\eea Using the
expression of the matrix elements in the above in terms of Schur
functions \bea \nn\langle \eta|\prod_{i}\Gamma_{+}(x_{i})|0\rangle=
\langle 0|\prod_{i}\Gamma_{-}(x_{i})|\eta\rangle=s_{\eta}({\bf x})
\eea we get \bea Z(N,M,L)=\sum_{\eta,\eta_{1}\leq N}s_{\eta}({\bf
x}^{+})\,s_{\eta}({\bf x}^{-})\,. \eea It is easy to see that the
above expression satisfies the limiting behavior of $Z(N)$: \bea
Z(0,M,L)=1\,,\,\,\,\,Z(\infty,M,L)=\prod_{L>m>0,0>m'>-M}(1-x^{+}_{m}x^{-}_{m'})^{-1}.\eea

Specializing to the $(q,t)$ parameters by using the previously
established map: \bea\nn
&&\{x^{+}_{m}|L>m>0\}=\{t,t^2,t^3,\cdots,t^{L}\}\\\nn
&&\{x^{-}_{m}|0>m>-M\}=\{1,q,q^2,\cdots,q^{M-1}\} \eea we get \bea
Z(N,M,L)=\sum_{\eta,\eta_{1}\leq
N}s_{\eta}(t,t^2,t^3,\cdots,t^{L})\,s_{\eta}(1,q,q^{2},\cdots,
q^{M-1})\,. \eea Using the identity \bea
s_{\eta}(1,q,q^2,\cdots,q^{M-1})=\,q^{n(\eta)}\prod_{s\in\eta}\frac{1-q^{M+j-i}}{1-q^{h(s)}},
\eea where $n(\eta)=\sum_{i}(i-1)\eta_{i}$ we get \bea
Z(N,M,L)=\sum_{\eta,\eta_{1}\leq
N}\,t^{|\eta|}\,(t\,q)^{n(\eta)}\,\prod_{s\in\eta}\frac{(1-t^{L+j-i})(1-q^{M+j-i})}{(1-t^{h(s)})(1-q^{h(s)})}.
 \eea
It is easy to see that if either of $L,M$ or $N$ is equal to $0$ then the above generating function reduces to $1$ as it should. However, it is easy to see that this crystal partition function is {\it not} related to the refined partition function of the closed topological vertex geometry $X_{C}$ \footnote{This geometry is actually the resolution of the singular geometry $\mathbb{C}^{3}/\mathbb{Z}_{2}\times \mathbb{Z}_{2}$.} in any simple way. The refined partition function for $X_{C}$ is given in Appendix A.  It is easy to see that if instead of putting a ceiling in the crystal model we calculate the crystal partition function by weighing the diagonal slice with $Q$ (as in the crystal model of $X_{0}$) we get the refined partition function of $X_{C}$,
\bea
Z_{crystal}&=&\sum_{\eta}(-Q)^{|\eta|}s_{\eta}(t^{\frac{1}{2}},t^{\frac{3}{2}},\cdots,t^{M-\frac{1}{2}})s_{\eta}(q^{\frac{1}{2}},q^{\frac{3}{2}},\cdots,q^{L-\frac{1}{2}})\\\nn
&=&\frac{Z_{\tiny vertex}(Q)}{Z_{\tiny vertex}(0)}\,.
\eea
Where $Z_{\tiny vertex}$ is given by Eq(\ref{cvz}) given in Appendix A.

\section{Adjoint Theory and Periodic Schur Process}

In this section, we will discuss the refined crystal model for the 5D $U(1)$ theory with
an adjoint hypermultiplet. We will also consider its 4D limit. In 4D this theory has properties similar to the ${\mathcal N}=4$ theory and is expected to be ultraviolet finite with partition function having modular properties. We will see that the combinatorics of the partition function of this theory is closely related to cylindric partitions \cite{borodin}.

\subsection{Geometric Engineering of $U(1)$ Theory with Adjoint Hypermultiplet}

We will denote by $X_{H}$ the geometry which gives rise to ${\cal N}=2$ abelian gauge theory with one adjoint hypermultiplet via compactification of type IIA. The gauge theory is obtained by taking a special limit which we will discuss later. M-theory compactification of $X_{H}$ gives rise to a ${\cal N}=1$ 5D theory. We will consider this 5D theory on $\mathbb{R}^{4}\times S^{1}$ with the radius of $S^{1}$ given by $\beta$. The partition function we will compute in the next section using topological vertex formalism is the partition function of the compactified 5D gauge theory.

The geometry $X_{H}$ (and its mirror) was studied in detail in \cite{Donagi:1995cf, Witten:1997sc, Intriligator:1997pq, Hollowood:2003cv} and is the total space of a rank two bundle over an elliptic curve. The rank two bundle is the trivial bundle twisted by a line bundle with first Chern class equal to the mass, $m$, of the adjoint hypermultiplet. For $m=0$ we ${\cal N}=4$ gauge theory as the $X_{H}$ in this case is simply $E\times \mathbb{C}^2$ where $E$ is an elliptic curve. \\

\begin{wrapfigure}{l}{3in}
\centering \epsfxsize=1in \epsffile{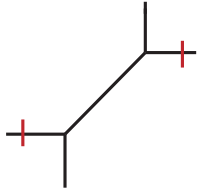} \caption{\small The "toric" diagram of $X_{H}$. There are two choices for the preferred direction required by the refined vertex: The internal $(1,1)$ line or the two vertical external lines.} \label{ppp}
\end{wrapfigure}
This geometry $X_{H}$ can also be obtained by partial compactfication of $X_{0}$ \cite{Hollowood:2003cv}. In terms of toric diagrams this corresponds to a non-planar toric diagram obtained from the toric diagram of $X_{0}$ by gluing two of the parallel external edges. This gives rise to another $\mathbb{P}^{1}$ in the geometry such that the new K\"ahler parameter is proportional to the mass $m$ of the adjoint hypermultiplet. This is shown in \figref{ppp}. In this case the refined vertex calculation can be done in two different ways corresponding to two different choices for the preferred direction: The internal $(1,1)$ line or the two vertical external edges.

\subsection{ Refined partition function}

We begin with the calculation of the refined partition function of this theory using the refined topological vertex formalism \cite{IKV} and the toric diagram of $X_{H}$ given above. Recall that the refined vertex calculation requires a preferred direction at each vertex such that all preferred directions of a given toric diagram be parallel\footnote{For an arbitrary toric diagram this may not be possible. An example is the toric diagram of ${\cal O}(-3)\mapsto \mathbb{P}^{2}$. This condition is actually equivalent to requiring that the corresponding CY3-fold be such that it gives rise to a supersymmetric gauge theory via geometric engineering. Thus this CY will be some $A_{n}$ fibration over a chain of $\mathbb{P}^{1}$'s. The preferred direction then corresponds to the base of this fibration.}. In the case of $X_{H}$ we see that there are two choices for the preferred direction corresponding to two seemingly different partition functions. Choosing the preferred direction to be the vertical external legs we get:
\bea\nn
\label{form1}
Z^{(1)}(Q,Q_{m},t,q)&=&\sum_{\lambda,\mu}(-Q_{m})^{|\mu|}(-Q)^{|\lambda|}C_{\lambda\mu\emptyset}(t,q)C_{\lambda^{t}\mu^{t}\emptyset}(q,t)\\
\nonumber
&=&\sum_{\lambda,\mu,\eta_{1},\eta_{2}}(-Q_{m})^{|\mu|}(-Q)^{|\lambda|}\left(\frac{q}{t}\right)^{\frac{|\eta_{1}|-|\eta_{2}|}{2}}s_{\lambda^{t}/\eta_{1}}(t^{-\rho})s_{\mu/\eta_{1}}(q^{-\rho})s_{\lambda/\eta_{2}}(q^{-\rho})s_{\mu^{t}/\eta_{2}}(t^{-\rho})\\
&=&M(t,q)^{-1}\sum_{\lambda,\mu}Q_{\bullet}^{|\lambda|}\,s_{\lambda/\mu}\Big(t^{-\rho+\frac{1}{2}}q^{-\frac{1}{2}},Q^{-1}t^{\rho}\Big)
s_{\lambda/\mu}\Big(q^{-\rho},\sqrt{\frac{q}{t}}Q_{m}^{-1}\,q^{\rho}\Big)\\\nn
&=&\prod_{i,j=1}^{\infty}(1-Q_{m}\,q^{i-\frac{1}{2}}\,t^{j-\frac{1}{2}})\,\widehat{Z}^{(1)}(Q,Q_{m},t,q)
\eea
where
\bea
\label{ins1}\nn
Q&=&e^{-T}\,,\,\,Q_{m}=e^{-T_{m}}\,,\,\,T,T_{m} \,\,\,\mbox{are the two K\"ahler parameters}\\\nn
M(t,q)&=&\prod_{i,j=1}^{\infty}(1-t^{i-1}\,q^{j})^{-1}\,,\,\,\,\,\,\,\,\,\,Q_{\bullet}=Q\,Q_{m}\\\
\widehat{Z}^{(1)}(Q,Q_{m},t,q)&=&\sum_{\lambda,\mu}Q^{|\lambda|}\,Q_{m}^{|\mu|}s_{\lambda/\mu}\Big(-Q_{m}^{(1)}\sqrt{\frac{q}{t}}\,q^{\rho},q^{-\rho}\Big)
s_{\lambda^{t}/\mu^{t}}\Big(Q_{m}^{(2)}\sqrt{\frac{t}{q}}\,t^{\rho},t^{-\rho}\Big)
\label{dfg}\eea
On the other hand choosing the internal $(1,1)$ leg to be the preferred direction we get:
\bea\nn
Z^{(2)}(Q,Q_{m},t,q)&=&\sum_{\lambda,\nu}(-Q)^{|\nu|}(-Q_m)^{|\lambda|}C_{\emptyset\lambda\nu}(t,q)C_{\emptyset\lambda^{t}\nu^{t}}(q,t)\\\nn
&=&\sum_{\nu}(-Q)^{|\nu|}q^{\frac{||\nu||^2}{2}}\,t^{\frac{||\nu^t||^2}{2}}\widetilde{Z}_{\nu}(t,q)\,\widetilde{Z}_{\nu^t}(q,t)
\prod_{i,j=1}^{\infty}(1-Q_{m}\,q^{-\mu_{i}-\rho_{j}}t^{-\mu^{t}_{j}-\rho_{i}})\\
&=&\prod_{i,j=1}^{+\infty}(1-Q_{m}\,t^{i-\frac{1}{2}}\,q^{j-\frac{1}{2}})\,\widehat{Z}^{(2)}(Q,Q_m,t,q)\,,
\label{form2}
\eea
where
\bea\label{form3}
\widehat{Z}^{(2)}(Q,Q_m,t,q)=\sum_{\nu}(QQ_{m})^{|\nu|}\prod_{s
\in\nu}
\frac{(1-Q_{m}\,t^{a(s)+\frac{1}{2}}\,q^{\ell(s)+\frac{1}{2}})(1-Q_{m}^{-1}\,q^{\ell
(s)+\frac{1}{2}}\,t^{a(s)+\frac{1}{2}})}{(1-t^{a(s)+1}\,q^{\ell(s)})(1-q^{\ell(s)+1}
\,t^{a(s)})}\,.
\eea
Note that:\\
$\blacksquare$ In Eq.(\ref{dfg}) we have introduced superscripts on $Q_{m}^{(1)}$ and $Q_{m}^{(2)}$ just to distinguish the arguments of the skew-Schur function from each other. But we will always take $Q_{m}^{(1)}=Q_{m}^{(2)}=Q_{m}$.\\
$\blacksquare$ In going from Eq. (\ref{form2}) to Eq. (\ref{form3}) we have used the following identity \cite{Nakajima:2003pg}:
\bea\nn
\frac{\prod_{i,j=1}^{\infty}\left(1-Q_{m}\,q^{-\nu_{i}-\rho_{j}}\,t^{-\nu_{j}^{t}-
\rho_{i}}\right)}{\prod_{i,j=1}^{\infty}\left(1-Q_{m}\,q^{-\rho_{i}}\,t^{-
\rho_{j}} \right)}=\prod_{s\in\nu}\left(1-Q_{m}\,q^{-\ell(s)-\frac{1}{2}}\,t^{-a
(s)-\frac{1}{2}}\right)\left(1-Q_{m}\, q^{\ell(s)+\frac{1}{2}}\,t^{a(s)+\frac{1}{2}}\right)
\eea

The two expressions $Z^{(1)}(Q,Q_m,t,q)$ and $Z^{(2)}(Q,Q_m,t,q)$ (and therefore $\widehat{Z}^{(1)}(Q,Q_m,t,q)$ and $\widehat{Z}^{(2)}(Q,Q_{m},t,q)$) appear different but are actually equal to each other as can be seen by expanding them in powers of $Q$ and $Q_m$ therefore from now on we will not use the superscript to distinguish them unless we need a specific form of the partition function. Thus we see that different choices for the preferred direction for a given toric diagram give rise to non-trivial $(q,t)$ identities involving ``principal specialization'' of the Macdonald function \cite{macdonald}. In section 4 we will give some other examples of $(q,t)$ identities arising from the refined topological vertex calculation.

In the gauge theory language $\widehat{Z}^{(1)}(Q,Q_m,t,q)$ or $\widehat{Z}^{(2)}(Q,Q_{m},t,q)$ is the contribution to the gauge theory partition function coming from instantons whereas the prefactor $\prod_{i,j}(1-Q_{m}t^{i-\frac{1}{2}}\,q^{j-\frac{1}{2}})$ is the perturbative contribution. For the moment we will ignore this prefactor and focus only on the instanton contribution. The partition function $Z(Q,Q_m,t,q)$ is invariant under the exchange of $Q$ and $Q_{m}$ but $\widehat{Z}(Q,Q_m,t,q)$ is not invariant under this exchange.

Using the identity
\bea
\sum_{\lambda,\mu}Q_{\bullet}^{|\lambda|}\,s_{\lambda/\mu}({\bf x})\,s_{\lambda/\mu}({\bf y})=
\prod_{k=1}^{\infty}\Big((1-Q_{\bullet}^{k})^{-1}\,\prod_{i,j=1}^{\infty}(1-Q_{\bullet}^{k}x_{i}y_{j})^{-1}\Big)\,,
\eea
we can write $Z(Q,Q_m,t,q)$ in Eq.(\ref{form1}) in a product form:
\bea\nn
Z(Q,Q_m,t,q)=M(t,q)^{-1}\prod_{k=1}^{\infty}\Big((1-Q_{\bullet}^k)\prod_{i,j=1}^{\infty}(1-Q_{\bullet}^{k}Q_{m}^{-1}q^{-i+\frac{1}{2}}t^{j-\frac{1}{2}})
(1-Q_{\bullet}^{k}Q^{-1}q^{i-\frac{1}{2}}t^{-j+\frac{1}{2}})\\\nn
\times(1-Q_{\bullet}^{k}\,q^{i-1}t^{j})
(1-Q_{\bullet}^{k-1}\,q^{-i+1}t^{-j})\Big)^{-1}.
\eea
If $|t|<1$ and $|q|<1$ (\textit{i.e.}, $\epsilon_{1}<0,\epsilon_{2}>0$) then we should write the above as
\bea\nn
Z(Q,Q_m,t,q)=\prod_{k=1}^{\infty}\Big((1-Q_{\bullet}^{k})^{-1}\prod_{i,j=1}^{\infty}\frac{(1-Q_{\bullet}^{k}\,Q_{m}^{-1}\,
q^{i-\frac{1}{2}}\,t^{j-\frac{1}{2}})(1-Q_{\bullet}^{k}\,Q^{-1}\,q^{i-\frac{1}{2}}t^{j-\frac{1}{2}})}
{(1-Q_{\bullet}^{k}\,q^{i-1}t^{j})(1-Q_{\bullet}^{k}\,q^{i}t^{j-1})}\Big)\,.
\eea
On the other hand if $|t|>1$ and $|q|<1$ (\textit{i.e.}, $\epsilon_{1}>0, \epsilon_{2}>0$) then we should write the above partition function in terms of $q$ and $t^{-1}$ so that
\bea\nn
Z(Q,Q_m,t,q)=\prod_{k=1}^{\infty}\Big((1-Q_{\bullet}^{k})^{-1}
\prod_{i,j}\frac{(1-Q_{\bullet}^{k}\,q^{i-1}t^{1-j})(1-Q_{\bullet}^{k}\,q^{i}t^{-j})}{(1-Q_{\bullet}^{k}Q_{m}^{-1}q^{i-\frac{1}{2}}\,t^{-j+\frac{1}{2}})(1-Q_{\bullet}^{k}
Q^{-1}
q^{i-\frac{1}{2}}t^{-j+\frac{1}{2}})}
\Big)\,.
\eea

\subsection{Field theory limit}
If we denote the mass of the adjoint by $m$ then the 4D field theory limit is given by\footnote{In identifying $Q_m$ with the mass $m$ we have introduced a factor of $\sqrt{\frac{t}{q}}$ so that the $m\mapsto 0$ limit gives the partition function of ${\mathcal N}=4$ gauge theory. For the discussion of  the relation between $(q,t)$ and the $\Omega$-background \cite{Nekrasov:2002qd} parameters $(\epsilon_1,\epsilon_2)$ we refer the reader to \cite{IKV}. }
\bea\label{4dftl}
Q_{m}=\sqrt{\frac{t}{q}}\,e^{-\beta m}\,,\,\,t=e^{\,\beta \epsilon_{1}}\,,q=e^{-\,\beta\epsilon_{2}}\,,\,\,\beta \mapsto 0\,.
\label{flimit}
\eea
We will see that the two instanton partition functions $\widehat{Z}^{(1)}(Q,Q_{m},t,q)$ and $\widehat{Z}^{(2)}(Q,Q_{m},t,q)$ give rise to an interesting identity in the above limit. We begin with $\widehat{Z}^{(2)}(Q,Q_{m},t,q)$ as it is easy to see what the field theory limit of this is:
\bea\nn
\widehat{Z}^{(2)}(Q,Q_m,t,q)\xmapsto{\beta\mapsto 0}&& \widehat{{\cal Z}}(Q,m,\epsilon_1,\epsilon_2)\\\nn
&=&\sum_{\nu}Q^{|\nu|}\prod_{s\in \nu}\frac{(a(s)+1+\vartheta\,\ell(s)-\widetilde{m})(a(s)+\vartheta\,(\ell(s)+1)+\widetilde{m})}
{(a(s)+1+\vartheta\,\ell(s))(a(s)+\vartheta(\ell(s)+1))},
\eea
where $\vartheta=-\epsilon_{2}/\epsilon_{1}$ and $\widetilde{m}=m/\epsilon_{1}$. The product inside the sum, in the expression above, gives a generalization of Nekrasov-Okounkov probability measure \cite{Nekrasov:2003rj} on the set of partitions \cite{kerov} \footnote{$h(i,j)=a(i,j)+\ell(i,j)+1$ is the hook length.},
\bea
\prod_{s\in \lambda}\frac{h(s)^{2}-\widetilde{m}^2}{h(s)^{2}}\mapsto \prod_{s\in \lambda}\frac{\Big(a(s)+1+\vartheta\,\ell(s)-\widetilde{m}\Big)\Big(a(s)+\vartheta\,\ell(s)+\vartheta+\widetilde{m}\Big)}
{\Big(a(s)+1+\vartheta\,\ell(s)\Big)\Big(a(s)+\vartheta\,\ell(s)+\vartheta)\Big)}\\\nn
\mapsto \prod_{s\in\lambda}
\frac{(1-Q_{m}\,t^{a(s)+1}\,q^{\ell(s)})(1-Q_{m}^{-1}\,q^{\ell(s)+1}\,t^{a(s)})}{(1-t^{a(s)+1}\,q^{\ell(s)})(1-q^{\ell(s)+1}\,t^{a(s)})}
\eea
It was shown in \cite{Nekrasov:2003rj} that for $\vartheta=1$ ($\epsilon_{1}+\epsilon_{2}=0$),
\bea
\widehat{{\cal Z}}(Q,m,\epsilon_1,-\epsilon_1)=\prod_{n=1}^{\infty}\Big(1-Q^n\Big)^{\widetilde{m}^2-1}.
\eea
Taking the field theory limit of $\widehat{Z}^{(1)}(Q,Q_{m},t,q)$ in Eq.(\ref{dfg}) is slightly more subtle. To take this limit first notice that the argument of the first skew-Schur function in Eq.(\ref{dfg}) is an infinite set which reduces to a finite set if we take $Q^{(1)}_{m}=\sqrt{\frac{t}{q}}\,q^{M}$ and analytically continue to $|q|<1$. The same is true for the argument of the second skew-Schur function:
\bea\label{ee1}
\{Q_{m}^{(1)}\sqrt{\frac{q}{t}}\,q^{\rho},q^{-\rho}\}\,\xmapsto{Q^{(1)}_{m}=\sqrt{\frac{t}{q}}\,q^{M}}&=&\{q^{\frac{1}{2}},q^{\frac{3}{2}},q^{\frac{5}{2}},\cdots,q^{M-\frac{1}{2}}\}\\\nn
\{Q_{m}^{(2)}\sqrt{\frac{t}{q}}\,t^{\rho},t^{-\rho}\}\,\xmapsto{Q^{(2)}_{m}=\sqrt{\frac{q}{t}}\,t^{L}}&=&\{t^{\frac{1}{2}},t^{\frac{3}{2}},t^{\frac{5}{2}},
\cdots,t^{L-\frac{1}{2}}\}\\\nn
\eea
With this identification Eq. (\ref{dfg}) gives
\bea\nn
\widehat{Z}^{(1)}(Q,Q_m,t,q)&=&\sum_{\lambda,\mu}(-Q)^{|\lambda|}\,(-Q_{m})^{|\mu|}
s_{\lambda/\mu}(q^{\frac{1}{2}},q^{\frac{3}{2}},q^{\frac{5}{2}},\cdots,q^{M-\frac{1}{2}})\,s_{\lambda^{t}/\mu^{t}}(t^{\frac{1}{2}},
t^{\frac{3}{2}},\cdots,t^{L-\frac{1}{2}})\\\nn
&=&\prod_{k=1}^{\infty}\Big((1-Q^{k}Q_{m}^{k})^{-1}\,\prod_{i,j=1}^{M,L}(1-Q^{k}Q_{m}^{k-1}\,q^{i-\frac{1}{2}}\,t^{j-\frac{1}{2}})\Big).
\eea
In the field theory limit given by Eq. (\ref{flimit}) we get\bea
\widehat{Z}^{(1)}(Q,Q_m,t,q)\xmapsto{\beta\mapsto 0}\widehat{\cal Z}(Q,m,\epsilon_1,\epsilon_2)=\prod_{k=1}^{\infty}(1-Q^k)^{ML-1}\,.
\eea
But since $Q_{m}^{(1)}=Q_{m}^{(2)}=\sqrt{\frac{t}{q}}\,e^{-\beta m}$,
\bea
\sqrt{\frac{t}{q}}\,q^{M}&=&\sqrt{\frac{q}{t}}\,t^{L}\,\Rightarrow\,
\epsilon_{1}(L-1)=-\epsilon_{2}(M-1) \Rightarrow L-1=\vartheta (M-1)\,,\\\nn
m&=&\epsilon_{2}\,M,\,.
\eea
which implies that
\bea
ML-1&=&(M-1)(\vartheta\,M+1)\\\nn
&=&-\frac{(m-\epsilon_{1})(m-\epsilon_{2})}{\epsilon_{1}\,\epsilon_{2}}\,\,.
\eea
Thus we see that in the field theory limit
\bea
\widehat{Z}(Q,Q_m,t,q)\xmapsto{\beta \mapsto 0}  \widehat{\cal Z}(Q,m,\epsilon_1,\epsilon_2)=\prod_{k=1}^{\infty}(1-Q^k)^{-\frac{(m-\epsilon_1)(m-\epsilon_2)}{\epsilon_1\,\epsilon_2}}.
\eea
This gives us the following interesting identity:
\bea\nn
\shadowbox{$\sum_{\nu}Q^{|\nu|}\prod_{s\in \nu}\frac{(a(s)+1+\vartheta\,\ell(s)-\tilde{m})(a(s)+\vartheta\,(\ell(s)+1)+\tilde{m})}
{(a(s)+1+\vartheta\,\ell(s))(a(s)+\vartheta(\ell(s)+1))}=
\prod_{k=1}^{\infty}(1-Q^k)^{\frac{(\tilde{m}-1)(\tilde{m}+\vartheta)}{\vartheta}}$}
\eea

\section{Periodic Schur Process}

Let us denote by $\textsf{P}$ the set of Young diagrams. A periodic Schur process is a random process defined on $\textsf{P}^{2K}$ such that it assigns to a set $\{\lambda^{(a)},\mu^{(a+1)}\,|\,a=0,1\cdots,K-1\}$ of $2K$ partitions the weight \cite{borodin}
\bea
\frac{1}{G_{K}}\times \varphi^{|\lambda^{(0)}|}\,\prod_{a=0}^{K-1}s_{\lambda^{(a)}/\mu^{(a+1)}}({\bf x}_{a+1})s_{\lambda^{(a+1)}/\mu^{(a+1)}}({\bf y}_{a+1})
\eea
where $\lambda^{(K)}=\lambda^{(0)}$, ${\bf x}_{a},{\bf y}_{a}$ are specializations of the algebra of symmetric functions and $G_{K}$ is the partition function of the process,
\bea
G_{K}(\varphi,{\bf x},{\bf y}):=\sum_{\lambda^{(0)}\mu^{(1)},\cdots,\lambda^{(K-1)},\mu^{(K)}}\varphi^{|\lambda^{(0)}|}\,\prod_{a=0}^{K-1}s_{\lambda^{(a)}/\mu^{(a+1)}}({\bf x}_{a+1})s_{\lambda^{(a+1)}/\mu^{(a+1)}}({\bf y}_{a+1})\,.
\eea
We will consider the case when $K=1$ which is closely related, as we will see, to the counting of cylindric plane partitions. For $K=1$ the weight assigned to the pair $\{\lambda,\mu\}$ is
\bea
\frac{1}{G_{1}}\times s_{\lambda/\mu}({\bf x})\,s_{\lambda/\mu}({\bf y})
\eea
and
\bea\label{Kequals1}
G_{1}(\varphi,{\bf x},{\bf y})=\sum_{\lambda,\mu}\varphi^{|\lambda|}\,s_{\lambda/\mu}({\bf x})\,s_{\lambda/\mu}({\bf y})\,.
\eea
If we take a particular specialization
\bea
{\bf x}&=&{\bf x}(t,q,Q)=\{t^{-\rho+\frac{1}{2}}\,q^{-\frac{1}{2}},Q^{-1}t^{\rho}\}\\
{\bf y}&=&{\bf y}(t,q,Q_{m})=\{q^{-\rho},\sqrt{\frac{q}{t}}\,Q_{m}^{-1}\,q^{\rho}\}\,,
\eea
and take $\varphi=Q_{\bullet}$ then from Eq(\ref{form1}) it follows that
\bea
G_{1}(Q_{\bullet},{\bf x}(t,q,Q),{\bf y}(t,q,Q_{m}))=M(t,q)\,Z^{(1)}(Q,Q_{m},t,q)
\eea
Thus the partition function of the $K=1$ periodic Schur process is precisely the partition function of the abelian gauge theory with an adjoint hypermultiplet. It was shown in \cite{borodin} that the $K=1$ Schur process is related to the counting of cylindric partitions.
The cylindric partitions, first introduced in \cite{GK}, are generalizations of the plane partitions. However, for our purposes, the reparameterization of them in \cite{borodin} is more suitable, which we largely follow.

A cylindric plane partition of type $(n,\ell)$ is an infinite array $\{\pi_{i,j}\,|\,i,j\in \mathbb{Z}\}$ of non-negative numbers such that:
\bea\nn
&&\pi_{i,j} \,\mbox{is weakly decreasing in both $i$ and $j$}\,,\\\nn
&&\pi_{i,j}=\pi_{i+n,j-\ell}\,.
\eea
The figure below shows an example of a cylindric partition. It is an infinite periodic diagram with one period shown between the vertical lines. Since the partitions on the vertical lines are identical we can glue them together and instead consider a finite diagram on a cylinder with period $n+\ell$.
\begin{figure}[h]\begin{center}
\includegraphics[width=5in]{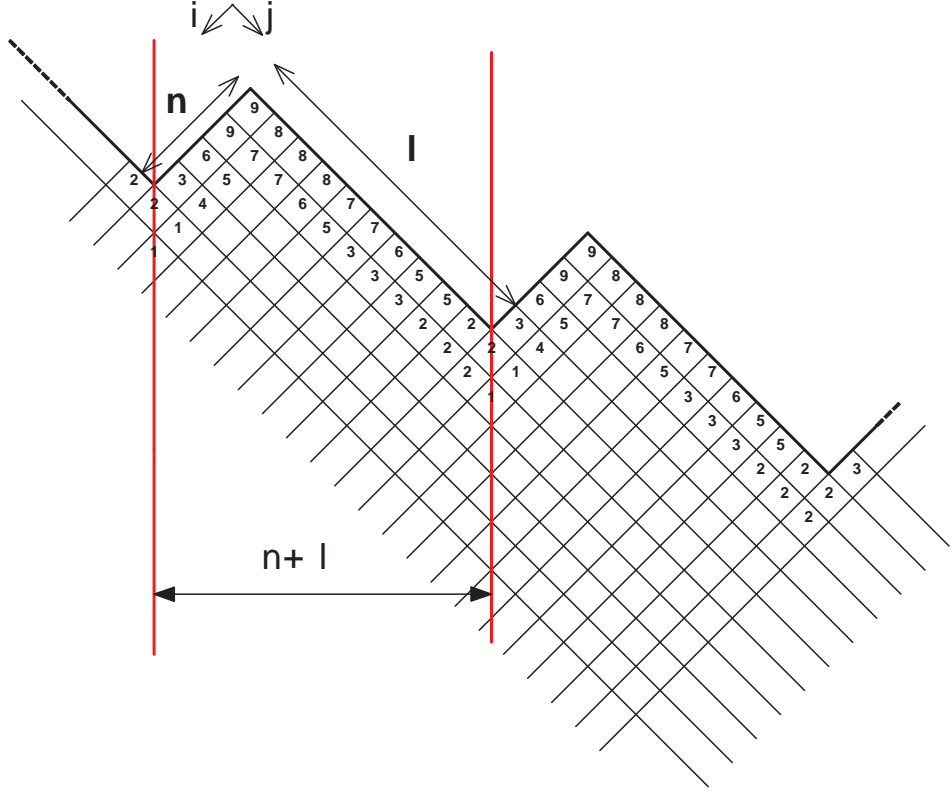}
\caption{\small An example for a cylindric partition with $(n,\ell)=(4,10)$.} \label{cyl}
\end{center}\end{figure}

Let us define $\mathbb{G}^{n,\ell}(s)$ to be the generating function of cylindric plane partitions of type $(n,\ell)$,
\bea
\mathbb{G}^{n,\ell}(s)=\sum_{\mbox{\tiny cylindric partitions $\pi$ of type $(n,\ell)$}}\,s^{|\pi|}\,,
\eea
where $|\pi|=\sum_{i=1,j=1}^{n,\ell}\pi_{i,j}$. This generating function was determined in \cite{borodin} and is given by
\bea\label{gf}
\mathbb{G}^{n,\ell}(s)=\prod_{k=1}^{\infty}\Big((1-s^{k\,(n+\ell)})^{-1}\prod_{i=1,j=1}^{\ell,n}(1-s^{k\,(n+\ell)-i-j+1})^{-1}\Big)\,.
\eea
We will show that this generating function is {\it exactly} the partition function of $X_{H}$ after an identification of parameters. To see this recall that the partition function of $X_{H}$ (Eq(\ref{form1})) is given by
\bea\label{eeee}
Z^{(1)}(Q,Q_m,t,q)&=&M(t,q)^{-1}\,\mathbb{Z}(Q,Q_{m},t,q)\\\nn
\mathbb{Z}(Q,Q_{m},t,q)&=&\sum_{\lambda,\mu}Q_{\bullet}^{|\lambda|}\,
s_{\lambda/\mu}\Big(t^{-\rho}\sqrt{\frac{t}{q}},Q^{-1}t^{\rho}\Big)
s_{\lambda/\mu}\Big(q^{-\rho},\sqrt{\frac{q}{t}}Q_{m}^{-1}\,q^{\rho}\Big)\,
\eea
The arguments of the two skew-Schur functions in the above equation are an infinite set of variables. By quantizing the two K\"ahler parameters (and analytic continuation to $|q|<1, |t|<1$) we can convert these infinite set of variables into a finite set:
\bea
\{t^{i}q^{-\frac{1}{2}},Q^{-1}t^{-i+\frac{1}{2}}\,|\,i\geq 1\} &&\xmapsto{Q=\sqrt{\frac{q}{t}}\,t^{-\ell}}\,\,\,\{t^{i}\,q^{-\frac{1}{2}}\,|\,i=1,2,\cdots, \ell\}\\\nn
\{q^{i-\frac{1}{2}},\sqrt{\frac{q}{t}}Q_{m}^{-1}\,q^{-i+\frac{1}{2}}\,|\,i\geq 1\} &&\xmapsto{Q_{m}=\sqrt{\frac{t}{q}}\,q^{-n}}\,\,\,\{q^{i-\frac{1}{2}}\,|\,i=1,2,\cdots,n\}
\eea
Eq(\ref{eeee}) becomes
\bea\nn
\mathbb{Z}\Big(Q=\sqrt{\frac{q}{t}}\,t^{-\ell},Q_m=\sqrt{\frac{t}{q}}\,q^{-n},t,q\Big)
&=&\sum_{\lambda,\mu}
(t^{-\ell}\,q^{-n})^{|\lambda|}\,
s_{\lambda/\mu}\Big(t\,q^{-\frac{1}{2}},t^{2}\,q^{-\frac{1}{2}},\cdots,t^{\ell}\,q^{-\frac{1}{2}}\Big)\\\nn
&\times&
s_{\lambda/\mu}\Big(q^{\frac{1}{2}},q^{\frac{3}{2}},\cdots,q^{n-\frac{1}{2}}\Big)\,\\\nn
&=&\prod_{k=1}^{\infty}\Big((1-(t^{-\ell}\,q^{-n})^k)\prod_{i,j=1}^{n,\ell}(1-q^{i-1-k\,n}\,t^{j-k\,\ell})
\eea
Comparing the above with Eq(\ref{gf}) we get
\bea\label{hj}\shadowbox{$
 \mathbb{Z}\Big(Q=s^{\ell},Q_m=s^{n},s^{-1},s^{-1}\Big)\,=\,\mathbb{G}^{n,\ell}(s)\,$}
\eea
The two K\"ahler parameters $T,T_{m}$ are quantized and given by the positive integers $(n,\ell)$ which define the type of the cylindric partitions:
\bea\nn
T&=&\epsilon_{1}\,\ell+\frac{\epsilon_{1}+\epsilon_{2}}{2}\\\nn
T_{m}&=&-\epsilon_{2}\,n-\frac{\epsilon_{1}+\epsilon_{2}}{2}
\eea

\section{ Hilbert schemes and Vertex Operators}

The Hilbert schemes of $\mathbb{C}^{2}$ ($\mbox{Hilb}^{\bullet}[\mathbb{C}^{2}])$ play a central role in the Nekrasov's derivation of the gauge theory partition functions using localization\cite{Nekrasov:2002qd}. In this section we see that the topological string partition function of $X_{H}$ can be written in terms of certain vertex operators which are generalization of operators related with the cohomology of the Hilbert scheme $\mbox{Hilb}^{\bullet}[\mathbb{C}^{2}]$ and were studied recently in \cite{carlsson}.

Recall that the instanton part of the gauge theory partition function of is given by Eq(\ref{ins1}),
\bea\label{p1}\nn
\widehat{Z}^{(1)}(Q,Q_{m},t,q)&=&\sum_{\lambda,\mu}Q^{|\lambda|}\,Q_{m}^{|\mu|}\,
s_{\lambda/\mu}\Big(-Q_{m}\sqrt{\frac{q}{t}}\,q^{\rho},q^{-\rho}\Big)
s_{\lambda^{t}/\mu^{t}}\Big(Q_{m}\sqrt{\frac{t}{q}}\,t^{\rho},\,t^{-\rho}\Big)\\
&=&\sum_{\lambda,\mu}Q_{\bullet}^{\lambda}\,
s_{\lambda/\mu}\Big(Q_{m}\sqrt{\frac{q}{t}}\,q^{\rho},q^{-\rho}\Big)
s_{\lambda/\mu}\Big(Q_{m}^{-1}\,t^{\rho},\sqrt{\frac{t}{q}}\,t^{-\rho}\Big)
\eea
Where $Q_{\bullet}=Q\,Q_{m}$ and  in going from the first line in the above equation to the second line we have used the properties of the principal specialization of the Schur functions:
\bea
s_{\lambda^{t}/\mu^{t}}(z\,q^{-\rho},q^{\rho})=(-z)^{|\lambda|-|\mu|}s_{\lambda/\mu}(z^{-1}q^{-\rho},q^{\rho})
\eea
The skew-Schur function in Eq(\ref{p1}) can be written as a matrix element of an operator. To see this note that
\bea
s_{\lambda/\mu}({\bf x})=\sum_{\eta}c^{\lambda}_{\eta\,\mu}\,s_{\eta}({\bf x})\,,
\eea
where $c^{\lambda}_{\eta\,\mu}$ are the Littlewood-Richardson coefficients. Let us define an operator $\widetilde{\alpha}_{\nu}$ labeled by a partition $\nu$ such that
\bea
\widetilde{\alpha}_{\nu}|\eta\rangle =\sum_{\lambda}c^{\lambda}_{\nu\,\eta}\,|\lambda\rangle\,,
\eea
It follows from the above that
\bea
c^{\lambda}_{\nu\,\eta}=\langle\lambda|\widetilde{\alpha}_{\nu}|\eta\rangle=\langle \eta|\widetilde{\alpha}_{\nu}^{\dagger}|\lambda\rangle\,.
\eea
Then we see that the skew-Schur function can be written as
\bea
s_{\lambda/\mu}({\bf x})&=&\sum_{\eta}c^{\lambda}_{\eta\,\mu}\,s_{\eta}({\bf x})=\sum_{\eta}\langle \lambda|\widetilde{\alpha}_{\eta}|\mu\rangle\,s_{\eta}({\bf x})\\\nn
&=&\langle \lambda|\Big(\sum_{\eta}\,s_{\eta}({\bf x})\widetilde{\alpha}_{\eta}\Big)|\mu\rangle\,=\langle \mu|\Big(\sum_{\eta}\,s_{\eta}({\bf x})\widetilde{\alpha}_{\eta}^{\dagger}\Big)|\lambda\rangle\,.
\eea
The operators $\widetilde{\alpha}_{\nu}$ are give by
\bea
\widetilde{\alpha}_{\nu}=\sum_{\mu}\,z_{\mu}^{-1}\,\chi^{\nu}(\mu)\,\alpha_{\mu}\,,
\eea
where $z_{\mu}=1^{m_{1}}m_{1}!2^{m_{2}}m_{2}!\cdots $ for a partition $\mu=(1^{m_{1}}2^{m_{2}}\cdots)$ and $\chi^{\mu}$ is the character of the symmetric group. The operator $\alpha_{\nu}=\alpha_{\nu_{1}}\alpha_{\nu_{2}}\cdots$ and $[\alpha_{n},\alpha_{m}]=n\delta_{n+m,0}$. It is easy to see that in terms of power sum symmetric functions $p_{n}({\bf x})=\sum_{i}x_{i}^{n}$\,\footnote{$p_{\mu}({\bf x})=p_{\mu_{1}}({\bf x})p_{\mu_{2}}({\bf x})\cdots=\sum_{\lambda}\chi^{\lambda}(\mu)\,s_{\lambda}({\bf x}), s_{\lambda}({\bf x})=\sum_{\mu}\,z_{\mu}^{-1}\,\chi^{\lambda}(\mu)\,p_{\mu}({\bf x})\,, \sum_{\lambda}\chi^{\lambda}(\mu)\chi^{\lambda}(\nu)=\delta_{\mu,\nu}\,, \sum_{\mu}z_{\mu}^{-1}\,\chi^{\lambda}(\mu)\chi^{\nu}(\mu)=\delta_{\lambda,\nu}\,.$}
\bea\\\nn
\mathbb{W}({\bf x})&=&\sum_{\eta}\,s_{\eta}({\bf x})\widetilde{\alpha}_{\eta}=\sum_{\mu}\,z_{\mu}^{-1}\,p_{\mu}({\bf x})\,\alpha_{\mu}\\
&=&\mbox{exp}\Big(\sum_{n=1}^{\infty}\frac{p_{n}({\bf x})}{n}\,\alpha_{n}\Big)
\eea
Using the above realization of the skew-Schur function as a matrix element we can write
\bea\nn
\sum_{\lambda,\mu}Q_{1}^{|\lambda|}\,Q_{2}^{|\mu|}\,s_{\lambda/\mu}({\bf x})\,s_{\lambda/\mu}({\bf y})&=&\sum_{\lambda,\mu}Q_{1}^{|\lambda|}Q_{2}^{|\mu|}\,\langle \lambda|\Big(\sum_{\eta}\,s_{\eta}({\bf x})\widetilde{\alpha}_{\eta}\Big)|\mu\rangle\,\langle \mu|\Big(\sum_{\eta}\,s_{\eta}({\bf y})\widetilde{\alpha}_{\eta}^{\dagger}\Big)|\lambda\rangle\,\\
&=&\mbox{Tr}\Big(Q_{1}^{H}\,\mathbb{W}({\bf x})\,Q_{2}^{H}\,\mathbb{W}({\bf y})^{\dagger}\Big)
\eea
Where $H\,|\lambda\rangle =|\lambda|\,|\lambda\rangle$.
Thus the instanton part of the gauge theory partition function can be written as the following trace:
\bea\nn
\widehat{Z}^{(1)}(Q,Q_{m},t,q)
&=&\mbox{Tr}\Big(Q_{\bullet}^{H}\mbox{exp}\Big(\sum_{n\geq 1}\frac{q^{\frac{n}{2}}(1-z_{1}^n)}{n(1-q^n)}\alpha_{n}\Big)
\,\mbox{exp}\Big(\sum_{n\geq 1}\frac{t^{n}(1-z_{2}^n)}{n\,q^{\frac{n}{2}}(1-t^n)}\alpha_{-n}\Big)\Big)\,,
\eea
where $z_{1}=Q_{m}\,\sqrt{\frac{q}{t}}$ and $z_{2}=Q_{m}^{-1}\,\sqrt{\frac{q}{t}}$. In the 4D field theory limit given by Eq(\ref{4dftl}) the above trace becomes:
\bea
\widehat{{\cal Z}}(Q,m,\epsilon_{1},\epsilon_{2})=\mbox{Tr}\Big(Q^{H}\,\mbox{exp}\Big(\frac{m}{\epsilon_{2}}\sum_{n>0}\frac{1}{n}\alpha_{n}\Big)
\mbox{exp}\Big(\frac{m-\epsilon_{1}-\epsilon_{2}}{\epsilon_{1}}
\sum_{n>0}\frac{1}{n}\alpha_{-n}\Big)\Big)
\eea
And specializing the Omega background $\epsilon_{1}=-\epsilon_{2}=0$ we get ($\widetilde{m}=\frac{m}{\epsilon_{1}}=-\frac{m}{\epsilon_{2}}$)
\bea
\widehat{\cal Z}(Q,m,-\epsilon_{2},\epsilon_{2})&=&\mbox{Tr}\Big(Q^{H}\,\mbox{exp}\Big(-\widetilde{m}\sum_{n>0}\frac{1}{n}\alpha_{n}\Big)
\mbox{exp}\Big(\widetilde{m}
\sum_{n>0}\frac{1}{n}\alpha_{-n}\Big)\Big)\\\nn
&=&\sum_{\lambda}Q^{|\lambda|}\,\langle\lambda|W_{\widetilde{m}}|\lambda\rangle
\eea
where
\bea
W_{\widetilde{m}}=\mbox{exp}\Big(-\widetilde{m}\,\sum_{n>0}\frac{1}{n}\alpha_{n}\Big)
\mbox{exp}\Big(\widetilde{m}
\sum_{n>0}\frac{1}{n}\alpha_{-n}\Big)
\eea
In \cite{carlsson} it was shown that the vertex operator $W_{\widetilde{m}}$ has matrix elements given by intersection over the Hilbert Scheme of $\mathbb{C}^{2}$,
\bea
\langle\mu|W_{\widetilde{m}}|\lambda\rangle=\int_{\mbox{Hilb}^{k}[\mathbb{C}^{2}]\times\mbox{Hilb}^{l}[\mathbb{C}^{2}]}\,\omega_{\lambda}\,\omega_{\mu}\,e(E)\,,
\label{int}
\eea
where $\omega_{\lambda}$ and $\omega_{\mu}$ are the pull-backs to the product of the cohomology classes of $\mbox{Hilb}^{k}[\mathbb{C}^{2}]$  and $\mbox{Hilb}^{l}[\mathbb{C}^{2}]$ respectively. $k$ and $l$ are given by $|\lambda|$ and $|\mu|$ and $E$ is a bundle on the product whose fiber at $(I,J)\in  \mbox{Hilb}^{k}[\mathbb{C}^{2}]\times\mbox{Hilb}^{l}[\mathbb{C}^{2}]$ is given by \cite{carlsson}
\bea
E|_{(I,J)}=\chi({\cal O}(m))-\chi(I,J\otimes {\cal O}(m))\,.
\eea
The proof of Eq(\ref{int}) was shown to be equivalent to the following identity \cite{carlsson}
\bea\nn
\langle E^{-\frac{\tilde{m}}{\theta}}\,(E^{\dagger})^{\tilde{m}+\theta-1}\,J_{\lambda},J_{\mu}\rangle_{\vartheta}=(-1)^{|\lambda|+|\mu|}\,\delta_{\lambda,\mu}\,
\prod_{s\in\lambda}(a(s)+\vartheta(\ell(s)+1)+\tilde{m})\,\prod_{s\in \mu}(a(s)+1+\ell(s)\,\vartheta-\tilde{m})\,.
\eea
Where $J_{\lambda}$ are the integral form of the Jack polynomials, $E=\mbox{exp}\Big(\sum_{n>0}\frac{(-1)^{n}}{n}p_{n}\,\Big)$ and $\langle \cdot,\cdot\rangle_{\vartheta}$ is the $\vartheta$ dependent product defined over the ring of symmetric functions. Using the above identity we can write the 4D gauge theory partition function as
\bea\nn
\sum_{\lambda}\varphi^{|\lambda|}\frac{\langle E^{-\frac{\tilde{m}}{\theta}}\,(E^{\dagger})^{\tilde{m}+\theta-1}\,J_{\lambda},J_{\lambda}\rangle_{\vartheta}}{\langle J_{\lambda},J_{\lambda}\rangle_{\vartheta}}&=&\sum_{\lambda}(-\varphi)^{|\lambda|}\prod_{s\in\lambda}
\frac{(a(s)+\vartheta(\ell(s)+1)+\tilde{m})(a(s)+1+\ell(s)\,\vartheta-\tilde{m})}{(a(s)+\vartheta(\ell(s)+1))(a(s)+1+\ell(s)\,\vartheta)}\\\nn
&&=\prod_{k=1}^{\infty}(1-\varphi^k)^{\tilde{m}^2-1}\,.
\eea
  Since the two parameter generalization of the Jack polynomials are Macdonald polynomials therefore it is not surprising that the 5D gauge theory partition function can be written using the $(q,t)$-dependent product $\langle \cdot\,,\,\cdot\rangle_{q,t}$ defined on the ring of symmetric functions.  This product is defined such that \cite{macdonald}
\bea\label{inn}
\langle P_{\lambda},P_{\mu}\rangle_{q,t}=\delta_{\lambda\,\mu}(-1)^{|\lambda|+|\mu|}\prod_{s\in\lambda}\frac{1-q^{\ell(s)+1}\,t^{a(s)}}{1-q^{\ell(s)}\,t^{a(s)+1}}\,.
\eea
The integral form of the Macdonald polynomials ${\cal J}_{\lambda}(t,q)$ is defined as \cite{Haiman}
\bea
{\cal J}_{\lambda}&=&\Big(\prod_{s\in\lambda}(1-q^{\ell(s)}\,t^{a(s)+1})\Big)P_{\lambda}\,,
\eea
such that
\bea
\langle {\cal J}_{\lambda},{\cal J}_{\mu}\rangle_{q,t}&=&\delta_{\lambda\,\mu}(-1)^{|\lambda|+|\mu|}\prod_{s\in\lambda}(1-q^{\ell(s)+1}\,t^{a(s)})(1-q^{\ell(s)}\,t^{a(s)+1})\,\\\nn
\lim_{t\mapsto 1}\frac{{\cal J}|_{q=t^{\theta}}}{(\mbox{ln}t)^{|\lambda|}}&=&J_{\lambda}\,.
\eea
In terms of this product we can write the 5D gauge theory partition function as
\bea\nn
\sum_{\lambda}\varphi^{|\lambda|}\frac{\langle {\cal G}(Q_{m},t,q)\,{\cal J}_{\lambda},{\cal J}_{\lambda}\rangle_{q,t}}{\langle {\cal J}_{\lambda},{\cal J}_{\lambda}\rangle_{q,t}}=\sum_{\nu}\varphi^{|\nu|}\prod_{s
\in\nu}
\frac{(1-Q_{m}\,t^{a(s)+\frac{1}{2}}\,q^{\ell(s)+\frac{1}{2}})(1-Q_{m}^{-1}\,q^{\ell
(s)+\frac{1}{2}}\,t^{a(s)+\frac{1}{2}})}{(1-t^{a(s)+1}\,q^{\ell(s)})(1-q^{\ell(s)+1}
\,t^{a(s)})}\,.
\eea
Where
\bea
{\cal G}(Q_{m},t,q)&=&E\Big(z_{1},1;q\Big)\,E\Big(\sqrt{\frac{t}{q}}z_{2},\sqrt{\frac{t}{q}};t\Big)^{\dagger}\\\nn
E\Big(x,y;q\Big)&=&\mbox{exp}\Big(\sum_{n\geq 1}\frac{(-1)^{n}}{n}(\frac{x^{n}-y^{n}}{q^{\frac{n}{2}}-q^{-\frac{n}{2}}})p_{n}\Big)\\\nn
z_{1}&=&Q_{m}\sqrt{\frac{q}{t}}\,,\,\,z_{2}=Q_{m}^{-1}\,\sqrt{\frac{q}{t}}\,.
\eea
The adjoint of the operator $E$ is defined with respect to the inner product Eq(\ref{inn}). With respect to this inner product if we take $p_{k}$ to be the operator which multiplies the function with $p_{k}$ then $p_{k}^{\dagger}$ is given by
\bea
p_{k}^{\dagger}=k\Big(\frac{1-t^{k}}{1-q^{k}}\Big)\frac{\partial}{\partial\,p_{k}}\,.
\eea
Using the fact that integral form of the Macdonald polynomials can be interpreted as equivariant K-homology classes on $\mbox{Hilb}^{\bullet}[\mathbb{C}^{2}]$ it is possible to realize these operators in terms of equivariant bundles over $\mbox{Hilb}^{\bullet}[\mathbb{C}^{2}]$ \cite{wip2}. The relation with cylindric partitions suggests that the integral in Eq(\ref{int}) counts the number of cylindric partitions of a certain shape.

\section{ Refined Topological Vertex and $(q,t)$ Identities}
In this section, we will use the refined topological vertex
formalism to obtain certain $(q,t)$ identities. Recall that in calculating the partition
function of a toric Calabi-Yau threefold using the refined vertex a
preferred edge has to be chosen for each vertex such that in the
toric diagram all the preferred edges are parallel \cite{IKV}. For a given
toric diagram there may be more than one such choice of the
preferred direction. In such a case the expression for the partition
function may look different for different choices of the preferred
direction giving rise to identities involving the K\"ahler
parameters of the Calabi-Yau threefold and the equivariant
parameters $q$ and $t$.

These identities can also be directly obtained using fiber-base
duality of ${\cal N}=2$ gauge theories \cite{Katz:1997eq} and
Nekrasov's instanton calculus. Given that the refined topological vertex
computation can only be done for geometries giving rise to gauge
theories (via geometric engineering), the slicing independence of the
refined vertex and the fiber-base duality are one and the same
thing. We do not provide a rigorous mathematical proof of these
identities. In each case we make use of a computer code; we expand
each partition function corresponding to a slicing at a certain
order in the K\"{a}hler parameters and compare term by term. The
simplest such identity is the famous summation identity which
relates a sum over Schur functions to a sum over Macdonald functions,
\bea\nn
\sum_{\nu}\varphi^{\ell(\nu)}\,s_{\nu}(q^{-\rho})\,s_{\nu^{t}}(t^{-\rho})=\prod_{i,j=1}^{\infty}(1-\varphi\,q^{i-\frac{1}{2}}\,t^{j-\frac{1}{2}})
=\sum_{\nu}\varphi^{\ell(\nu)}
P_{\nu}(q^{-\rho};t,q)\,P_{\nu^{t}}(t^{-\rho};q,t). \eea

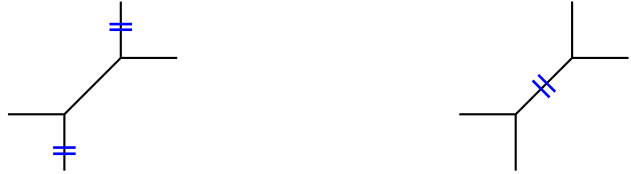
\begin{figure}[h]
\begin{center} $\begin{array}{c@{\hspace{1in}}c}
\multicolumn{1}{l}{\mbox{}} &
    \multicolumn{1}{l}{\mbox{}} \\ [-0.53cm]
{

\begin{pspicture}(4,2)(6,3.2)
\psline[unit=0.75cm,linecolor=black](4,3)(5,4)
\psline[unit=0.75cm,linecolor=black](3,3)(4,3)
\psline[unit=0.75cm,linecolor=black](4,2)(4,3)
\psline[unit=0.75cm,linecolor=black](5,4)(5,5)
\psline[unit=0.75cm,linecolor=black](5,4)(6,4)

\psline[unit=0.75cm,linecolor=black](12,3)(13,4)
\psline[unit=0.75cm,linecolor=black](11,3)(12,3)
\psline[unit=0.75cm,linecolor=black](12,2)(12,3)
\psline[unit=0.75cm,linecolor=black](13,4)(13,5)
\psline[unit=0.75cm,linecolor=black](13,4)(14,4)

\psline[unit=0.75cm,linecolor=blue,linestyle=solid,linewidth=1pt](12.7,3.4)(12.4,3.7)
\psline[unit=0.75cm,linecolor=blue,linestyle=solid,linewidth=1pt](12.6,3.3)(12.3,3.6)

\psline[unit=0.75cm,linecolor=blue,linestyle=solid,linewidth=1pt](4.8,4.5)(5.2,4.5)
\psline[unit=0.75cm,linecolor=blue,linestyle=solid,linewidth=1pt](4.8,4.6)(5.2,4.6)

\psline[unit=0.75cm,linecolor=blue,linestyle=solid,linewidth=1pt](3.8,2.3)(4.2,2.3)
\psline[unit=0.75cm,linecolor=blue,linestyle=solid,linewidth=1pt](3.8,2.41)(4.2,2.41)
\end{pspicture}}
\\ [0.0cm] \mbox{} & \mbox{}
\end{array}$
\caption{\small Two choices of the preferred direction labelled by the blue
lines.} \label{toric}\end{center}
\end{figure}

The toric diagram which gives rise to this identity is shown in \figref{toric}.\\

\subsection{ Identities}

Here we want to demonstrate five explicit examples of identities
using the slicing independence of the refined topological vertex.

For {\it almost} all toric geometries we generically have three
distinct choices for the preferred direction. For every internal
edge along the preferred direction there is a sum over all
partitions and for every internal edge not along the preferred
direction the sum can be performed explicitly to give infinite
products. However, for each of the three choices of the preferred
direction we do not always get a distinct expression. It is possible
that two different choices lead indeed to the same expressions with K\"{a}hler
classes and the corresponding labels for the partitions are
appropriately exchanged. We demonstrate examples for this case in
the following section.

\subsubsection{\sc Example 1: $X_{0}:={\cal O}(-1)\oplus {\cal O}(-1)\mapsto \mathbb{P}^{1}$}

Our first example is the toric geometry $X_{0}$. We can use the refined topological
vertex to determine the refined partition function. The toric
diagram and the two possible choices for the preferred direction are shown in gluing of the refined vertex are shown in
\figref{toric}. The refined partition function for the choice of the preferred direction shown in \figref{toric}(b) is given
by\bea
Z(t,q,Q)&:=&\sum_{\nu}(-Q)^{|\nu|}\,C_{\emptyset\,\emptyset\,\nu}(t,q)\,\,C_{\emptyset\,\emptyset\,\nu^{t}}(q,t)\\\nn
&=&\sum_{\nu}\frac{Q^{|\nu|}(-1)^{|\nu|}\,q^{\frac{\Arrowvert\nu\Arrowvert^{2}}{2}}\,t^{\frac{\Arrowvert\nu^{t}\Arrowvert^{2}}{2}}}
{\prod_{s\in
\nu}(1-t^{a(s)+1}q^{\ell(s)})(1-t^{a(s)}q^{\ell(s)+1})}\,.\eea
A different representation of the partition function can be obtained by choosing the preferred
directions as shown in \figref{toric}(a). The refined partition function with this choice is given by \bea\nn
Z(t,q,Q)&=&\sum_{\lambda}(-Q)^{|\lambda|}\,C_{\lambda\,\emptyset\,\emptyset}(t,q)\,
C_{\lambda^{t}\,\emptyset\,\emptyset}(q,t)= \sum_{\lambda}(-Q)^{|\lambda|}s_{\lambda^{t}}(t^{-\rho})\,s_{\lambda^{t}}(q^{-\rho})\\\nn
&=&\prod_{i,j=1}^{\infty}(1-Q\,q^{i-\frac{1}{2}}\,t^{j-\frac{1}{2}})=\mbox{Exp}\left \{
-\sum_{n=1}^{\infty}\frac{Q^{n}}{n(q^{\frac{n}{2}}-q^{-\frac{n}{2}})(t^{\frac{n}{2}}-t^{-\frac{n}{2}})}\right
\}\,. \eea Identifying the above two representations of the
partition function we get the following identity \bea
\sum_{\nu}\frac{Q^{|\nu|}(-1)^{|\nu|}\,q^{\frac{\Arrowvert\nu\Arrowvert^{2}}{2}}\,t^{\frac{\Arrowvert\nu^{t}\Arrowvert^{2}}{2}}}
{\prod_{s\in
\nu}(1-t^{a(s)+1}q^{\ell(s)})(1-t^{a(s)}q^{\ell(s)+1})}=\mbox{Exp}\left\{-\sum_{n=1}^{\infty}\frac{Q^{n}}{n(q^{\frac{n}{2}}-q^{-\frac{n}{2}})(t^{\frac{n}{2}}-t^{-\frac{n}{2}})}\right
\} \eea which is a specialization of the identity Eq. (5.4) of
\cite{macdonald} and was also derived in \cite{nakajima-2003}.

\subsubsection{\sc Example 2: ${\cal O}(0)\oplus {\cal
O}(-2)\mapsto \mathbb{P}^{1}$} This geometry can be obtained from
local $\mathbb{P}^{1}\times \mathbb{P}^{1}$ by taking the size of
one of the $\mathbb{P}^{1}$ very large and is the resolution of $\mathbb{C}\times \mathbb{C}^{2}/\mathbb{Z}_{2}$.
The toric geometry and the two possible choices for the preferred direction are shown in the \figref{twochoices} below.

\begin{figure}[h]\begin{center}
$\begin{array}{c@{\hspace{1in}}c} \multicolumn{1}{l}{\mbox{}} &
    \multicolumn{1}{l}{\mbox{}} \\ [-0.53cm]
{
\begin{pspicture}(0,0)(4,4)

\psline[unit=0.5cm,linewidth=1pt,linecolor=black](0,2)(0,3.5)
\psline[unit=0.5cm,linewidth=1pt,linecolor=black](0,3.3)(0,5)
\psline[unit=0.5cm,linewidth=1pt,linecolor=black](0,2)(1.2,2)
\psline[unit=0.5cm,linewidth=1pt,linecolor=black](1,2)(2,2)
\psline[unit=0.5cm,linewidth=1pt,linecolor=black](0,5)(1.6,5)
\psline[unit=0.5cm,linewidth=1pt,linecolor=black](1,5)(2,5)
\psline[unit=0.5cm,linewidth=1pt,linecolor=black](0,2)(-1,1)
\psline[unit=0.5cm,linewidth=1pt,linecolor=black](-2,0)(-0.7,1.3)
\psline[unit=0.5cm,linewidth=1pt,linecolor=black](0,5)(-1,6)
\psline[unit=0.5cm,linewidth=1pt,linecolor=black](-0.8,5.8)(-2,7)

\psline[unit=0.5cm,linewidth=1pt,linecolor=black](12,2)(12,3.5)
\psline[unit=0.5cm,linewidth=1pt,linecolor=black](12,3.3)(12,5)
\psline[unit=0.5cm,linewidth=1pt,linecolor=black](12,2)(13.2,2)
\psline[unit=0.5cm,linewidth=1pt,linecolor=black](13,2)(14,2)
\psline[unit=0.5cm,linewidth=1pt,linecolor=black](12,5)(13.6,5)
\psline[unit=0.5cm,linewidth=1pt,linecolor=black](13,5)(14,5)
\psline[unit=0.5cm,linewidth=1pt,linecolor=black](12,2)(11,1)
\psline[unit=0.5cm,linewidth=1pt,linecolor=black](10,0)(11.3,1.3)
\psline[unit=0.5cm,linewidth=1pt,linecolor=black](12,5)(11,6)
\psline[unit=0.5cm,linewidth=1pt,linecolor=black](11.2,5.8)(10,7)

\psline[unit=0.5cm,linewidth=1pt,linecolor=blue](-0.3,3.5)(0.3,3.5)
\psline[unit=0.5cm,linewidth=1pt,linecolor=blue](-0.3,3.3)(0.3,3.3)

\psline[unit=0.5cm,linewidth=1pt,linecolor=blue](13,4.7)(13,5.3)
\psline[unit=0.5cm,linewidth=1pt,linecolor=blue](13.2,4.7)(13.2,5.3)
\psline[unit=0.5cm,linewidth=1pt,linecolor=blue](13,1.7)(13,2.3)
\psline[unit=0.5cm,linewidth=1pt,linecolor=blue](13.2,1.7)(13.2,2.3)
\put(0,-0.5){a)}

\put(6,-0.5){b)}

\end{pspicture}}
\\ [-0.2cm] \mbox{} & \mbox{}
\end{array}$
\caption{\small Two possible choices for the preferred direction, the
internal line (a) and the parallel external lines (b).}
\label{twochoices}\end{center}
\end{figure}
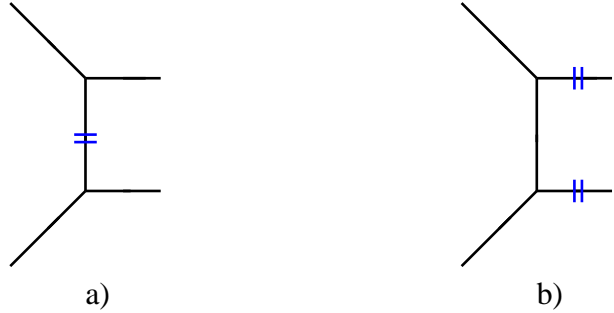
The refined partition function for the case \figref{twochoices}(a) is given by \bea
Z(t,q,Q)&=&\sum_{\nu}Q^{|\nu|}(-1)^{|\nu|}\,C_{\emptyset\,\emptyset\,\nu}(t,q)\,f_{\nu}(t,q)\,C_{\emptyset\,\emptyset\,\nu^{t}}(q,t)\\\nn
&=&\sum_{\nu}(-Q)^{|\nu|}\widetilde{Z}_{\nu}(t,q)\widetilde{Z}_{\nu^{t}}(q,t)\,q^{\frac{\Arrowvert\nu\Arrowvert^{2}}{2}}\,t^{\frac{\Arrowvert\nu^{t}\Arrowvert^{2}}{2}}\,f_{\nu}(t,q)\\\nn
&=&\sum_{\nu}\frac{(Q\sqrt{\tfrac{q}{t}})^{|\nu|}\,t^{\Arrowvert\nu^{t}\Arrowvert^{2}}}{\prod_{s\in
\nu}(1-t^{a(s)+1}\,q^{\ell(s)})(1-t^{a(s)}\,q^{\ell(s)+1})}\,.\eea
The partition function for case (b) of \figref{twochoices} is given
by, \bea
Z(t,q,Q)&=&\sum_{\lambda}Q^{|\lambda|}(-1)^{|\lambda|}\,C_{\emptyset\,\lambda\,\emptyset}(t,q)\,f_{\lambda}(t,q)\,\,
C_{\lambda^{t}\,\emptyset\,\emptyset}(t,q)\,\\\nn &=&
\sum_{\lambda}(-Q)^{|\lambda|}\Big(\frac{q}{t}\Big)^{\frac{\Arrowvert\lambda\Arrowvert^2}{2}}
t^{\frac{\kappa(\lambda)}{2}}\,s_{\lambda^{t}}(q^{-\rho})\,f_{\lambda}(t,q)\,s_{\lambda}(t^{-\rho})\,\,\,\\\nn
&=&\sum_{\lambda}(Q\sqrt{\tfrac{q}{t}})^{|\lambda|}\,s_{\lambda^{t}}(t^{-\rho})\,s_{\lambda^{t}}(q^{-\rho})
=\prod_{i,j=1}^{\infty}\Big(1-Q\,q^{i}\,t^{j-1}\Big)^{-1}\\\nn
&=&\,\mbox{Exp}\left \{
\sum_{n=1}^{\infty}\frac{Q^{n}\Big(\frac{q}{t}\Big)^{\frac{n}{2}}}{n(q^{\frac{n}{2}}-q^{-\frac{n}{2}})(t^{\frac{n}{2}}-t^{-\frac{n}{2}})}\right
\}\,. \eea
Thus we get the identity (after rescaling $Q$ and interchanging $q$ and $t$)
\bea
\sum_{\nu}\frac{Q^{|\nu|}\,q^{\Arrowvert\nu^{t}\Arrowvert^{2}}}{\prod_{s\in
\nu}(1-q^{a(s)+1}\,t^{\ell(s)})(1-q^{a(s)}\,t^{\ell(s)+1})}=\mbox{Exp}\left \{
\sum_{n=1}^{\infty}\frac{Q^{n}\Big(\frac{q}{t}\Big)^{\frac{n}{2}}}{n(q^{\frac{n}{2}}-q^{-\frac{n}{2}})(t^{\frac{n}{2}}-t^{-\frac{n}{2}})}\right
\}\,.
\eea

\subsubsection{\sc Example 3}
Our third example is that of the geometry giving rise to 5D $U(1)$
gauge theory with a single adjoint. The toric diagram of this
geometry is shown in \figref{U(1)5Dad} below.
\begin{figure}[h]
\begin{center}
$\begin{array}{c@{\hspace{1in}}c} \multicolumn{1}{l}{\mbox{}} &
    \multicolumn{1}{l}{\mbox{}} \\ [-0.4cm]
 \includegraphics[width=1.2in]{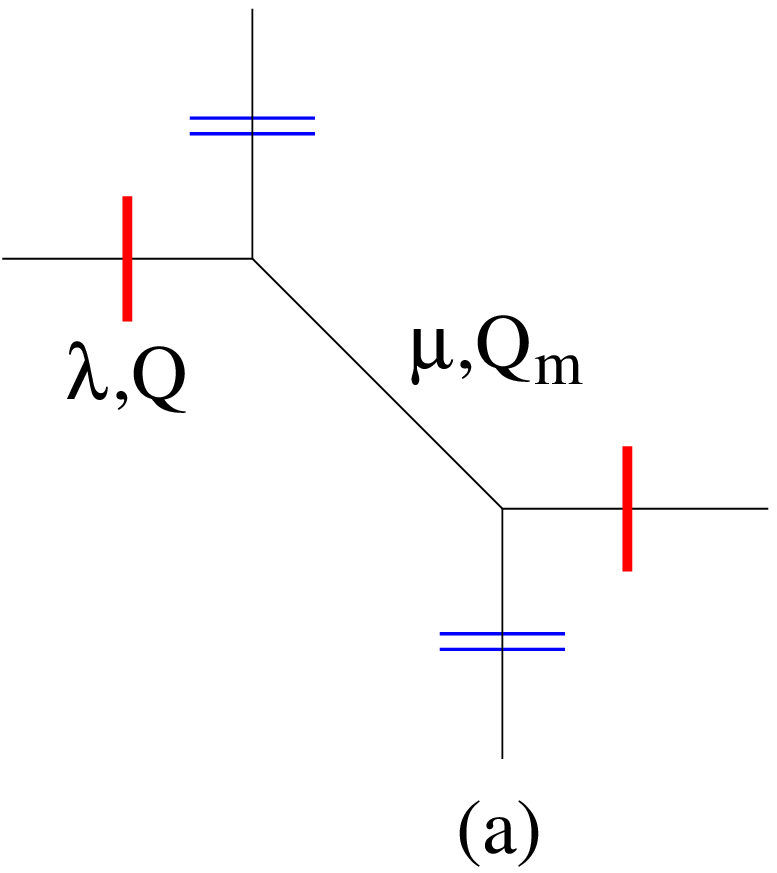} &
\includegraphics[width=1.2in]{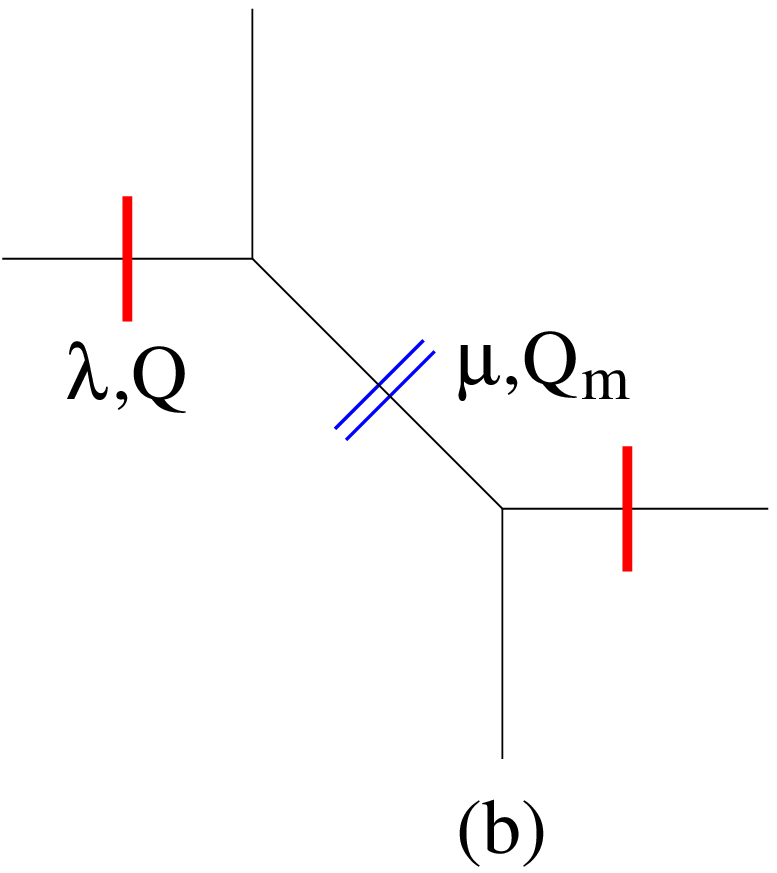}\\

\\ [0.4cm]
\end{array}$
\caption{\small{The toric diagram for the geometry giving rise to 5D
supersymmetric $U(1)$ theory with adjoint matter.}}
\label{U(1)5Dad}
\end{center}
\end{figure}
\figref{U(1)5Dad}(a) and \figref{U(1)5Dad}(b) show two
possible choices for the preferred direction needed for refined
vertex calculation. The single thick line indicates the gluing of
the corresponding edges giving rise to non-planar diagrams. In
\figref{U(1)5Dad}(a) the preferred direction is along the
non-compact edges whereas in \figref{U(1)5Dad}(b) the preferred
direction is along one of the compact edge.

The refined vertex calculation for \figref{U(1)5Dad}(a) gives:
\begin{eqnarray}
Z_{(a)}&=&\sum_{\lambda,\mu}(-Q_{m})^{|\mu|}(-Q)^{|\lambda|}C_{\lambda\mu\emptyset}(t,q)C_{\lambda^{t}\mu^{t}\emptyset}(q,t)\\
\nonumber
&=&\sum_{\lambda,\mu,\eta_{1},\eta_{2}}(-Q_{m})^{|\mu|}(-Q)^{|\lambda|}\left(\frac{q}{t}\right)^{\frac{|\eta_{1}|-|\eta_{2}|}{2}}s_{\lambda^{t}/\eta_{1}}(t^{-\rho})s_{\mu/\eta_{1}}(q^{-\rho})s_{\lambda/\eta_{2}}(q^{-\rho})s_{\mu^{t}/\eta_{2}}(t^{-\rho})\\
\nonumber
&=&\prod_{i',j'=1}^{\infty}(1-Q_{m}q^{-\rho_{i'}}t^{-\rho_{j'}})\prod_{k=1}^{\infty}
\Big[(1-Q^{k}Q_{m}^{k})^{-1}\prod_{i,j=1}^{\infty}(1-Q^{k}Q_{m}^{k-1}q^{-\rho_{i}}t^{-\rho_{j}})\\\nn
&&(1-Q^{k}Q_{m}^{k}q^{\rho_{i}-1/2}t^{-\rho_{j}+1/2})
(1-Q^{k}Q_{m}^{k}q^{-\rho_{i}+1/2}t^{\rho_{j}-1/2})(1-Q^{k}Q_{m}^{k+1}q^{\rho_{i}}t^{\rho_{j}})\Big]\,.
\end{eqnarray}

Changing the preferred direction changes the expression of the
refined vertex calculation and therefore \figref{U(1)5Dad}(b)
gives:
\begin{eqnarray}
Z_{(b)}&=&\sum_{\lambda,\mu}(-Q_{m})^{|\mu|}(-Q)^{|\lambda|}C_{\emptyset\lambda\mu}(t,q)C_{\emptyset\lambda^{t}\mu^{t}}(q,t)\\
\nonumber
&=&\sum_{\mu}(-Q_{m})^{|\mu|}\left(\frac{q}{t}\right)^{\frac{\Arrowvert\mu\Arrowvert^{2}-\Arrowvert\mu^{t}\Arrowvert^{2}}{2}}P_{\mu^{t}}(t^{-\rho};q,t)P_{\mu}(q^{-\rho};t,q)\prod_{i,j=1}^{\infty}(1-Qq^{-\mu_{i}-\rho_{j}}t^{-\mu^{t}_{j}-\rho_{i}})
\end{eqnarray}

\subsubsection{\sc Example 4}

The toric diagram of the geometry which gives rise to 5D U(1) gauge
theory with two hypermultiplets in the fundamental representation is
given in \figref{FIG:U3} below.

\begin{figure}[h]\begin{center}
\includegraphics[width=4.5in]{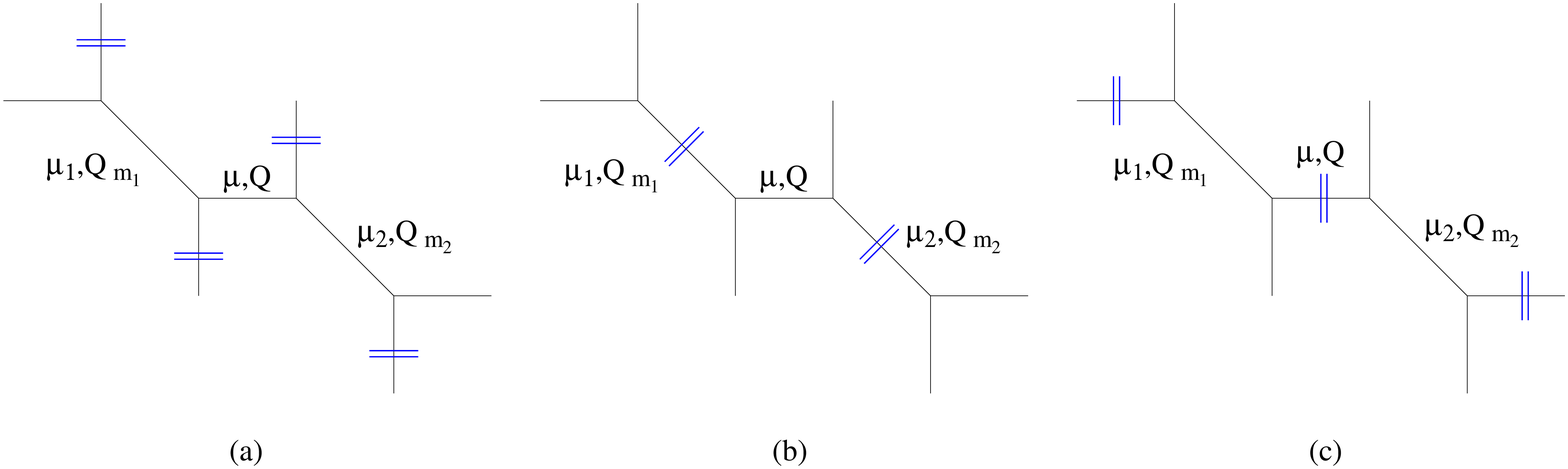}
\caption{\small{The toric diagram for the 5D supersymmetric $U(1)$
theory with $N_{f}=2$. The three distinct choices of the preferred
direction lead to three distinct expressions, with a) no sums over
partitions, with b) two sums over partitions and with c) one sum
over partitions.}} \label{FIG:U3}
\end{center}\end{figure}

The choice of the preferred direction is indicated by the short
double line. In this case there are three different choices for the
preferred direction giving rise to three seemingly different
expressions for the refined partition function.

\figref{FIG:U3}(a) gives:
\begin{eqnarray}
Z_{(a)}&=&
\sum_{\mu,\mu_{1},\mu_{2}}(-Q)^{|\mu|}(-Q_{m_{1}})^{|\mu_{1}|}(-Q_{m_{2}})^{|\mu_{2}|}C_{\emptyset\mu_{1}\emptyset}(t,q)C_{\mu\mu_{1}^{t}\emptyset}(q,t)C_{\mu^{t}\mu_{2}\emptyset}(t,q)C_{\emptyset\mu_{2}^{t}\emptyset}(q,t)\\
\nonumber
&=&\sum_{\mu,\mu_{1},\mu_{2},\eta_{1},\eta_{2}}(-Q)^{|\mu|}(-Q_{m_{1}})^{|\mu_{1}|}(-Q_{m_{2}})^{|\mu_{2}|}\left(\frac{t}{q}\right)^{\frac{|\eta_{1}|-|\eta_{2}|}{2}}s_{\mu_{1}}(q^{-\rho})s_{\mu^{t}_{1}/\eta_{1}}(t^{-\rho})s_{\mu^{t}/\eta_{1}}(q^{-\rho})\\
\nonumber &\times&
s_{\mu/\eta_{2}}(t^{-\rho})s_{\mu_{2}/\eta_{2}}(q^{-\rho})s_{\mu_{2}^{t}}(t^{-\rho})\\
\nonumber &=& \prod_{i,j=1}^{\infty}\frac{(1-Q
q^{-\rho_{i}}t^{-\rho_{j}})(1-Q_{m_{1}}
q^{-\rho_{i}}t^{-\rho_{j}})(1-Q_{m_2{}}
q^{-\rho_{i}}t^{-\rho_{j}})(1-QQ_{m_{1}}Q_{m_{2}}
q^{-\rho_{i}}t^{-\rho_{j}})}{(1-QQ_{m_{1}}
q^{-\rho_{i}-1/2}t^{-\rho_{j}+1/2})(1-QQ_{m_{2}}
q^{-\rho_{i}+1/2}t^{-\rho_{j}-1/2})}
\end{eqnarray}

\figref{FIG:U3}(b) gives:
\begin{eqnarray}\nn
Z_{(b)}&=&\sum_{\mu,\mu_{1},\mu_{2}}(-Q)^{|\mu|}(-Q_{m_{1}})^{|\mu_{1}|}(-Q_{m_{2}})^{|\mu_{2}|}C_{\emptyset\emptyset\mu_{1}}(t,q)C_{\emptyset\mu\mu_{1}^{t}}(q,t)C_{\emptyset\mu^{t}\mu_{2}}(t,q)C_{\emptyset\emptyset\mu_{2}^{t}}(q,t)\\
\nonumber
&=&\sum_{\mu_{1},\mu_{2}}(-Q_{m_{1}})^{|\mu_{1}|}(-Q_{m_{2}})^{|\mu_{2}|}\left(\frac{q}{t}\right)^{\frac{\Arrowvert\mu_{1}\Arrowvert^{2}-\Arrowvert\mu_{1}^{t}\Arrowvert^{2}+\Arrowvert\mu_{2}\Arrowvert^{2}-\Arrowvert\mu_{2}^{t}\Arrowvert^{2}}{2}}
P_{\mu_{1}^{t}}(t^{-\rho};q,t)P_{\mu_{1}}(q^{-\rho};t,q)P_{\mu_{2}^{t}}(t^{-\rho};q,t)\\
\nonumber &\times&
P_{\mu_{2}}(q^{-\rho};t,q)\prod_{i,j=1}^{\infty}(1-Qq^{-\mu_{1,i}-\rho_{j}}t^{-\mu_{2,j}^{t}-\rho_{i}})
\end{eqnarray}

\figref{FIG:U3}(c) gives:
\begin{eqnarray}\nn
Z_{(c)}&=&\sum_{\mu,\mu_{1},\mu_{2}}(-Q)^{|\mu|}(-Q_{m_{1}})^{|\mu_{1}|}(-Q_{m_{2}})^{|\mu_{2}|}C_{\mu_{1}\emptyset\emptyset}(t,q)C_{\mu_{1}^{t}\emptyset\mu}(q,t)C_{\mu_{2}\emptyset\mu^{t}}(t,q)C_{\mu_{2}^{t}\emptyset\emptyset}(q,t)\\
\nonumber
&=&\sum_{\mu}(-Q)^{|\mu|}\left(\frac{t}{q}\right)^{\frac{\Arrowvert\mu\Arrowvert^{2}-\Arrowvert\mu^{t}\Arrowvert^{2}}{2}}P_{\mu^{t}}(q^{-\rho};t,q)
P_{\mu}(t^{-\rho};q,t)\\\nonumber
&\times&\prod_{i,j=1}^{\infty}(1-Q_{m_{1}}q^{-\rho_{i}}t^{-\mu_{i}-\rho_{j}})(1-Q_{m_{2}}q^{-\mu^{t}_{i}-\rho_{j}}t^{-\rho_{i}})
\end{eqnarray}

{\bf \sc Flop transition}

\begin{figure}[h]\begin{center}
\includegraphics[width=4.5in]{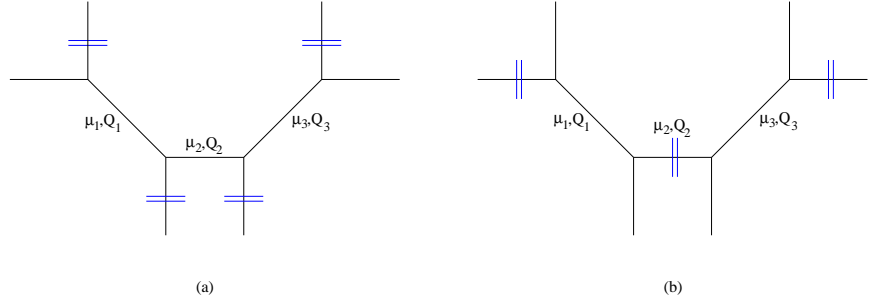}
\caption{\small The flop transition relates this geometry with that of
\figref{FIG:U3}. The corresponding geometrically engineered gauge
theory has masses of hypermultiplets of different signs.} \label{U}
\end{center}\end{figure}

In this case there are only two different choices for the preferred
direction.

\figref{U}(a) gives:
\begin{align}\nn
&Z_{(a)}=\sum_{\mu_{1},\mu_{2},\mu_{3}}(-Q_{1})^{|\mu_{1}|}(-Q_{2})^{|
\mu_{2}|}(-Q_{3})^{|\mu_{3}|}C_{\mu_{3}\emptyset\emptyset}(q,t)C_{\mu_
{3}^{t}\mu_{2}\emptyset}(t,q)f_{\mu_{2}}(t,q)C_{\mu_{2}^{t}\mu_{1}
\emptyset}(t,q)C_{\emptyset\mu_{1}^{t}\emptyset}(q,t)\\ \nonumber
&=\prod_{i,j=1}^{\infty}\frac{(1-Q_{1}q^{-\rho_{i}}t^{-\rho_{j}})(1-Q_
{3}q^{-\rho_{i}}t^{-\rho_{j}})(1-Q_{1}Q_{2}q^{-\rho_{i}+1}t^{-\rho_
{j}-1})(1-Q_{2}Q_{3}q^{-\rho_{i}+1}t^{-\rho_{j}-1})}{(1-Q_{2}q^{-\rho_
{i}+1/2}t^{-\rho_{j}-1/2})(1-Q_{1}Q_{2}Q_{3}q^{-\rho_{i}+3/2}t^{-\rho_
{j}-3/2})}
\end{align}

\figref{U}(b) gives:
\begin{align}\nn
&Z_{(b)}=\sum_{\mu_{1},\mu_{2},\mu_{3}}(-Q_{1})^{|\mu_{1}|}(-Q_{2})^{|
\mu_{2}|}(-Q_{3})^{|\mu_{3}|}C_{\emptyset\mu_{3}\emptyset}(q,t)C_
{\emptyset\mu_{3}^{t}\mu_{2}}(t,q)f_{\mu_{2}}(t,q)C_{\mu_{1}\emptyset
\mu_{2}^{t}}(q,t)C_{\mu_{1}^{t}\emptyset\emptyset}(t,q)\\ \nonumber
=&\sum_{\mu_{2}}(-{\widetilde Q_{2}}q^{1/2}t^{-1/2})^{|\mu_{2}|} t^
{\Arrowvert\mu_{2}^{t}\Arrowvert^{2}}t^{-\Arrowvert\mu_{2}\Arrowvert^
{2}/2}P_{\mu_{2}^{t}}(t^{-\rho};q,t)q^{-\Arrowvert\mu_{2}^{t}
\Arrowvert^{2}/2}P_{\mu_{2}}(q^{-\rho};t,q)\prod_{i,j=1}^{\infty}(1-Q_
{1}q^{-\rho_{i}}t^{-\mu_{2,i}^{t}-\rho_{j}})\\ \nonumber
&\times(1-Q_ {3}q^{-\rho_{i}}t^{-\mu_{2,i}^{t}-\rho_{j}})
\end{align}

\subsubsection{\sc Example 5}

Our final example is of the geometry giving rise to a quiver gauge
theory with gauge group $U(1)\times U(1)$. This theory, its
generalization with $U(k)^N$ gauge group and the corresponding
geometries were studied in \cite{Hollowood:2003cv} to which we refer
the reader for more details.

\begin{figure}[h]\begin{center}
\includegraphics[width=4in]{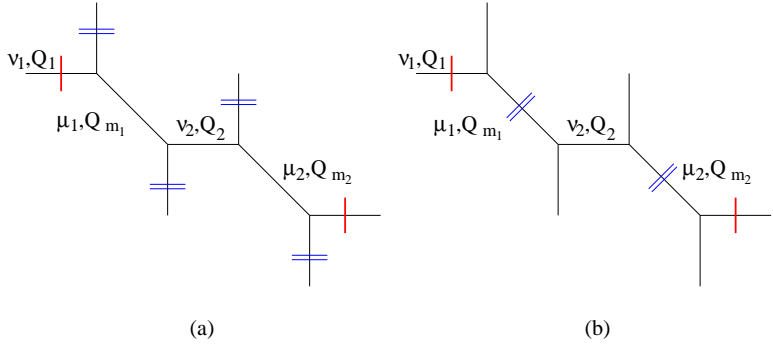}
\caption{\small{The toric diagram for the 5D ${\hat A}_{0}$ quiver
theory. The three distinct choices of the preferred directions are
not all distinct, the other choice, picking the preferred direction
along $\nu$'s will have basically the same form as the one in b).} }
\label{UU}
\end{center}\end{figure}

\figref{UU}(a) gives:
\begin{align}\nonumber
&Z_{(a)}=\sum_{\mu_{1},\mu_{2},\nu_{1},\nu_{2}}(-Q_{m_{1}})^{|\mu_{1}|}(-Q_{m_{2}})^{|\mu_{2}|}(-Q_{1})^{|\nu_{1}|}(-Q_{2})^{|\nu_{2}|}C_{\nu_{1}\mu_{1}\emptyset}(t,q)C_{\nu_{2}\mu_{1}^{t}\emptyset}(q,t)C_{\nu_{2}^{t}\mu_{2}\emptyset}(t,q)C_{\nu_{1}^{t}\mu_{2}^{t}\emptyset}(q,t)\\
\nonumber &=\sum_{\begin{array}{c}\mu_{1},\mu_{2},\nu_{1},\nu_{2},\\
\eta_{1},\eta_{2},\eta_{3},\eta_{4}\end{array}}(-Q_{m_{1}})^{|\mu_{1}|}(-Q_{m_{2}})^{|\mu_{2}|}(-Q_{1})^{|\nu_{1}|}(-Q_{2})^{|\nu_{2}|}\left(\frac{q}{t}\right)^{\frac{|\eta_{1}|-|\eta_{2}|+|\eta_{3}|-|\eta_{4}|}{2}}s_{\nu_{1}^{t}/\eta_{1}}(t^{-\rho})s_{\mu_{1}/\eta_{1}}(q^{-\rho})\\
\nonumber
&s_{\nu_{2}^{t}/\eta_{2}}(q^{-\rho})s_{\mu_{1}^{t}/\eta_{2}}(t^{-\rho})s_{\nu_{2}/\eta_{3}}(t^{-\rho})s_{\mu_{2}/\eta_{3}}(q^{-\rho})s_{\nu_{1}/\eta_{4}}(q^{-\rho})s_{\mu_{2}^{t}/\eta_{4}}(t^{-\rho})\\
\nonumber
&=\prod_{i,j=1}^{\infty}\frac{(1-Q_{1}q^{-\rho_{i}}t^{-\rho_{j}})(1-Q_{2}q^{-\rho_{i}}t^{-\rho_{j}})(1-Q_{m_{1}}Q_{2}q^{\rho_{i}-1/2}t^{-\rho_{j}+1/2})(1-Q_{m_{1}}Q_{1}Q_{2}q^{-\rho_{i}}t^{-\rho_{j}})}{(1-Q_{m_{1}}q^{\rho_{i}}t^{-\rho_{j}})(1-Q_{m_{1}}Q_{1}q^{-\rho_{i}+1/2}t^{-\rho_{j}-1/2})}\\
\nonumber
&\times\sum_{\mu_{2},\lambda}(-Q_{m_{2}})^{|\mu_{2}|}(-Q_{m_{1}}Q_{1}Q_{2})^{|\lambda|}s_{\mu_{2}^{t}/\lambda}(q^{\rho},Q_{2}q^{-\rho+1/2}t^{-1/2},Q_{m_{1}}Q_{2}q^{\rho},Q_{m_{1}}Q_{1}Q_{2}q^{-\rho+1/2}t^{-1/2})
\\ \nonumber &\times
s_{\mu_{2}/\lambda^{t}}(t^{\rho},Q_{1}t^{-\rho+1/2}q^{-1/2},Q_{m_{1}}Q_{1}t^{\rho},Q_{m_{1}}Q_{1}Q_{2}t^{-\rho+1/2}q^{-1/2})
\end{align}

\figref{UU}(b) gives:
\begin{align}\nonumber
&Z_{(b)}=\sum_{\mu_{1},\mu_{2},\nu_{1},\nu_{2}}(-Q_{m_{1}})^{|\mu_{1}|}(-Q_{m_{2}})^{|\mu_{2}|}(-Q_{1})^{|\nu_{1}|}(-Q_{2})^{|\nu_{2}|}C_{\emptyset\nu_{1}\mu_{1}}(t,q)C_{\emptyset\nu_{2}\mu_{1}^{t}}(q,t)C_{\emptyset\nu_{2}^{t}\mu_{2}}(t,q)C_{\emptyset\nu_{1}^{t}\mu_{2}^{t}}(q,t)
\\ \nonumber
&=\sum_{\mu_{1},\mu_{2}}(-Q_{m_{1}})^{|\mu_{1}|}(-Q_{m_{2}})^{|\mu_{2}|}(-Q_{1})^{|\nu_{1}|}(-Q_{2})^{|\nu_{2}|}\left(\frac{q}{t}\right)^{\frac{\Arrowvert\mu_{1}\Arrowvert^{2}-\Arrowvert\mu_{1}^{t}\Arrowvert^{2}+\Arrowvert\mu_{2}\Arrowvert^{2}-\Arrowvert\mu_{2}^{t}\Arrowvert^{2}}{2}}P_{\mu_{1}^{t}}(t^{-\rho};q,t)P_{\mu_{1}}(q^{-\rho};t,q)\\
\nonumber &\times
P_{\mu_{2}^{t}}(t^{-\rho};q,t)P_{\mu_{2}}(q^{-\rho};t,q)\prod_{i,j}^{\infty}(1-Q_{1}q^{-\mu_{2,i}-\rho_{j}}t^{-\mu_{1,j}^{t}-\rho_{i}})(1-Q_{2}q^{-\mu_{1,i}-\rho_{j}}t^{-\mu_{2,j}^{t}-\rho_{i}})
\end{align}

\section*{ Acknowledgments}
AI and CK would like to thank Charles Doran for many
valuable discussions.

\section{ Appendix A}
In this section we want to sketch the refined topological vertex computations for the resolved conifold, double-$\mathbb{P}^{1}$ and the closed refined topological vertex.
\subsubsection*{\sc Resolved conifold}
\begin{floatingfigure}[r]{2.0in}
\centering \epsfxsize=2.0in \epsffile{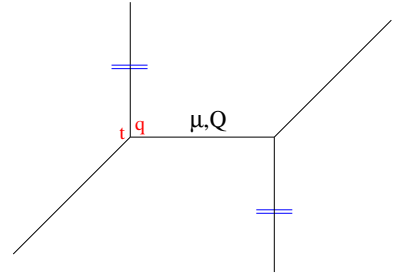} \caption{\small Toric diagram of resolved conifold} \label{conifold}
\end{floatingfigure}
\figref{conifold} shows the toric diagram of resolved conifold and our choice of $\{t,q\}$-parametrization. The double blue lines again show the preferred direction of the refined topological vertex. The partition function is given by
\bea
Z_{\tiny \mbox{vertex}}&=&\sum_{\mu}(-Q)^{|\mu|} C_{\emptyset\mu\emptyset}(t,q)C_{\emptyset\mu^{t}\emptyset}(q,t)\\ \nn
&=&\sum_{\mu}(-Q)^{|\mu|}s_{\mu}(q^{-\rho})s_{\mu^{t}}(t^{-\rho})\\ \nn
&=&\prod_{i=1}^{\infty}\prod_{j=1}^{\infty}\left(1-Q t^{i-1/2}q^{j-1/2}\right)
\eea
\subsubsection*{\sc Double-$\mathbb{P}^{1}$}
\begin{wrapfigure}{r}{2.0in}
\centering \epsfxsize=2.0in \epsffile{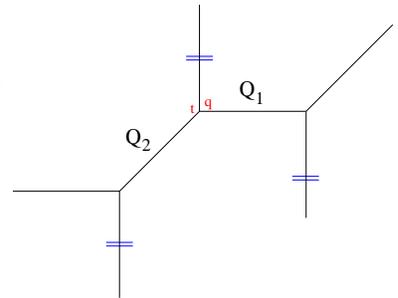} \caption{\small Toric diagram of double-$\mathbb{P}^{1}$}\label{double}
\end{wrapfigure}
\figref{double} shows the toric diagram of double-$\mathbb{P}^{1}$. The partition function reads
\begin{align}\nn
&Z_{\tiny \mbox{vertex}}=\sum_{\mu_{1},\mu_{2}}(-Q_{1})^{|\mu_{1}|}(-Q_{2})^{|\mu_{2}|}C_{\emptyset\mu_{1}\emptyset}(q,t)C_{\mu_{2}\mu_{1}^{t}\emptyset}(t,q)C_{\mu_{2}^{t}\emptyset\emptyset}(q,t)\\
&=\sum_{\mu_{1},\mu_{2},\eta}(-Q_{1})^{|\mu_{1}|}(-Q_{2})^{|\mu_{2}|}\left(\frac{q}{t}\right)^{|\eta|/2}s_{\mu_{1}}(t^{-\rho})s_{\mu_{1}^{t}/\eta}(q^{-\rho})\\ \nn
&\times s_{\mu_{2}}(q^{-\rho})s_{\mu_{2}^{t}/\eta}(t^{-\rho})\\ \nn
&=\prod_{i=1}^{\infty}\prod_{j=1}^{\infty}\frac{\left(1-Q_{1}t^{i-1/2}q^{j-1/2}\right)\left(1-Q_{2}t^{i-1/2}q^{j-1/2}\right)}{\left(1-Q_{1}Q_{2}t^{i-1}q^{j}\right)}
\end{align}
\subsubsection*{\sc Closed refined topological vertex}

\begin{figure}
\begin{center}
  \includegraphics[width=2in]{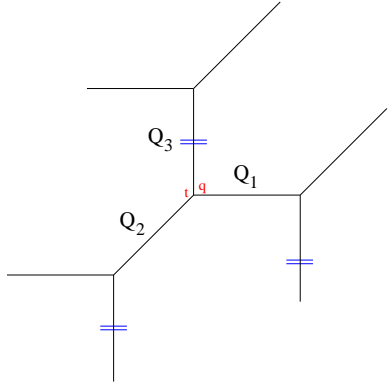}\\
  \caption{Toric diagram of closed refined topological vertex}\label{clos}
  \end{center}
\end{figure}

\figref{clos} shows the toric diagram of closed refined topological vertex. The partition function reads
\bea\nn
Z_{\tiny \mbox{vertex}}(Q_{3})&:=&\sum_{\mu_{1},\mu_{2},\mu_{3}}(-Q_{1})^{|\mu_{1}|}(-Q_{2})^{|\mu_{2}|}(-Q_{3})^{|\mu_{3}|}C_{\emptyset\emptyset\mu_{3}}(q,t)C_{\mu_{2}\mu_{1}^{t}\mu_{3}^{t}}(t,q)C_{\mu_{2}^{t}\emptyset\emptyset}(q,t)C_{\emptyset\mu_{1}\emptyset}(q,t)\\ \nn
&=&\sum_{\mu_{1},\mu_{2},\mu_{3},\eta}(-Q_{1})^{|\mu_{1}|}(-Q_{2})^{|\mu_{2}|}(-Q_{3})^{|\mu_{3}|}\left(\frac{q}{t}\right)^{|\eta|/2} \left(\frac{t}{q}\right)^{\Arrowvert\mu_{3}\Arrowvert^{2}/2}\\ \nn
&&\times P_{\mu_{3}^{t}}(q^{-\rho};t,q)\left(\frac{q}{t}\right)^{\Arrowvert\mu_{3}^{t}\Arrowvert^{2}/2}P_{\mu_{3}}(t^{-\rho};q,t)s_{\mu_{1}}(t^{-\rho}) s_{\mu_{1}^{t}/\eta}(t^{-\mu_{3}}q^{-\rho})\\ \nn
&&\times s_{\mu_{2}}(q^{-\rho})s_{\mu_{2}^{t}/\eta}(t^{-\rho}q^{-\mu_{3}^{t}})\\ \nn
&=&\sum_{\mu_{3}}(-Q_{3})^{|\mu_{3}|} \left(\frac{t}{q}\right)^{\frac{\Arrowvert\mu_{3}\Arrowvert^{2}-\Arrowvert\mu_{3}^{t}\Arrowvert^{2}}{2}}P_{\mu_{3}^{t}}(q^{-\rho};t,q)P_{\mu_{3}}(t^{-\rho};q,t)\\
&&\times\prod_{i=1}^{\infty}\prod_{j=1}^{\infty} \frac{(1-Q_{1}t^{-\mu_{3,i}+j-1/2}q^{j-1/2})(1-Q_{2}q^{-\mu_{3,i}^{t}+j-1/2}t^{i-1/2})}{(1-Q_{1}Q_{2}t^{i-1}q^{j})}
\label{cvz}
\eea

\section{ Appendix B: Useful Identities}

In this section, we want to give a short list of identities \cite{macdonald} or definitions which we have used in our computations. We also include a short proof of generalizing one of the identities. \par{The Schur functions define a basis for the symmetric function and the skew Schur functions $s_{\lambda/\mu}(x_{1},x_{2},\mathellipsis)$ have the following nice representation as a sum over all semi-standard Young tableau $\lambda/\mu$:}
\bea
s_{\lambda/\mu}(x_{1},x_{2},\mathellipsis)=\sum_{T_{\lambda/\mu}}x_{1}^{m_{1}}x_{2}^{m_{2}}\mathellipsis
\eea
where $m_{i}$ is the degeneracy of $i$ in the tableau. The Macdonald function $P_{\nu}(t^{-\rho};q,t)$ is defined by
\bea
P_{\nu}(t^{-\rho};q,t)=t^{\Arrowvert\nu^{t}\Arrowvert/2}{\widetilde Z_{\nu^{t}}}(t,q),
\eea
where $\rho_{i}=1/2-i$ and
\bea
{\widetilde Z_{\nu}}(t,q)=\prod_{s\in\nu}\frac{1}{1-t^{a(s)+1}q^{\ell(s)}}.
\eea
with $a(s)=\nu_{j}^{t}-i$ and $\ell(s)=\nu_{i}-j$ being the arm and leg length of $s=(i,j)\in\nu$, respectively.
The Schur functions have the following properties:
\bea
s_{\lambda/\mu}(x^{(1)},\mathellipsis,x^{(n)})&=&\sum_{(\nu)}\prod_{i=1}^{n}s_{\nu^{(i)}/\nu^{(i-1)}}(x^{(i)}),
\eea
where we have defined $\nu^{(0)}\equiv\mu$ and $\nu^{(n)}\equiv\lambda$. Note that $\nu^{(i-1)}\prec\nu^{(i)}$.
 \bea
 s_{\lambda/\mu}(q^{-\rho})=s_{\lambda^{t}/\mu^{t}}(-q^{\rho})
 \eea
 We extensively made use of the following sums
\bea
\sum_{\eta}s_{\eta/\nu}(x)s_{\eta/\mu}(y)&=&\prod_{i,j=1}^{\infty}(1-x_{i}y_{j})^{-1}\sum_{\tau}s_{\mu/\tau}(x)s_{\nu/\tau}(y),\\
\sum_{\eta}s_{\eta^{t}/\nu}(x)s_{\eta/\mu}(y)&=&\prod_{i,j=1}^{\infty}(1+x_{i}y_{j})\sum_{\tau}s_{\mu^{t}/\tau}(x)s_{\nu^{t}/\tau^{t}}(y),
\eea
where the right hand side of the equations reduce to the product if $\mu$ or $\nu$ is equal to the empty partition, since $s_{\emptyset/\tau}=1$ for $\tau=\emptyset$ and vanishes for any other $\tau$. The following product forms are known for the case of $\mu=\nu$ in the above identities and we also sum the left hand side over $\mu$
\bea\label{eq}
\sum_{\rho,\lambda}
q^{|\rho|}s_{\rho/\lambda}(x)s_{\rho/\lambda}(y)&=&\prod_{k=1}^{\infty}\left(1-q^{k}\right)^{-1}\prod_{i,j=1}^{\infty}\left(1-q^{k}x_{i}y_{j}\right)^{-1}\\
\sum_{\rho,\lambda}q^{|\rho|}s_{\rho^{t}/\lambda}(x)s_{\rho/\lambda^{t}}(y)&=&\prod_{k=1}^{\infty}\left(1-q^{k}\right)^{-1}\prod_{i,j=1}^{\infty}\left(1+q^{k}x_{i}y_{j}\right).
\eea
These last two identities are the ones we wan to generalize in the following way\footnote{We pick one of these two similar forms, the other form is obvious and the proof is identical.}:
\begin{align}\nonumber
&\sum_{\rho,\lambda}q^{|\rho|}s_{\rho/\lambda}(x^{(1)},x^{(2)},\mathellipsis,x^{(N)})s_{\rho/\lambda}(y^{(1)},y^{(2)},\mathellipsis,y^{(N)})=\prod_{k=1}^{\infty}\left(1-q^{k}\right)^{-1}\prod_{i,j=1}^{\infty}\left(1-q^{k}x_{i}^{(1)}y_{j}^{(1)}\right)^{-1}\\
\nonumber
&\prod_{i,j=1}^{\infty}\left(1-q^{k}x_{i}^{(2)}y_{j}^{(1)}\right)^{-1}\mathellipsis\prod_{i,j=1}^{\infty}\left(1-q^{k}x_{i}^{(1)}y_{j}^{(2)}\right)^{-1}\prod_{i,j=1}^{\infty}\left(1-q^{k}x_{i}^{(2)}y_{j}^{(2)}\right)^{-1}\mathellipsis\prod_{i,j=1}^{\infty}\left(1-q^{k}x_{i}^{(N)}y_{j}^{(M)}\right)^{-1}
\end{align}
where each of $x^{(j)}$ is an infinite series of variables
$(x^{(j)}_{1},x^{(j)}_{2},x^{(j)}_{3},\mathellipsis)$.

Using the fact that Schur functions are symmetric functions we can
introduce the following variables and make use of the Eq.\eqref{eq}:
\begin{eqnarray}\nonumber
w_{i}&=&x^{(i~\mbox{mod N})}_{\lfloor i/N \rfloor}\\ \nonumber
z_{i}&=&y^{(i~\mbox{mod M})}_{\lfloor i/M \rfloor},
\end{eqnarray}
hence,
\begin{equation}\nonumber
\sum_{\rho,\lambda}q^{|\rho|}s_{\rho/\lambda}(x^{(1)},x^{(2)},\mathellipsis,x^{(N)})s_{\rho/\lambda}(y^{(1)},y^{(2)},\mathellipsis,y^{(N)})=\sum_{\rho,\lambda}
q^{|\rho|}s_{\rho/\lambda}(w)s_{\rho/\lambda}(z)
\end{equation}
In these new variables the product will take the form
\begin{align}\nonumber
&\prod_{i,j=1}^{\infty}\left(1-q^{k}w_{i}z_{j}\right)^{-1}=\prod_{i=\{1,N+1,2N+1,\mathellipsis\}}\prod_{i=\{2,N+2,2N+2,\mathellipsis\}}\mathellipsis\prod_{j=1}^{\infty}\left(1-q^{k}w_{i}z_{j}\right)^{-1}\\
\nonumber
&=\prod_{i=\{1,N+1,2N+1,\mathellipsis\}}\prod_{j=1}^{\infty}\left(1-q^{k}w_{i}z_{j}\right)^{-1}\prod_{i=\{2,N+2,2N+2,\mathellipsis\}}\prod_{j=1}^{\infty}\left(1-q^{k}w_{i}z_{j}\right)^{-1}\mathellipsis
\\ \nonumber
&=\prod_{i=1}^{\infty}\prod_{j=1}^{\infty}\left(1-q^{k}x^{(1)}_{i}z_{j}\right)^{-1}\prod_{i=1}^{\infty}\prod_{j=1}^{\infty}\left(1-q^{k}x^{(2)}_{i}z_{j}\right)^{-1}\mathellipsis
\nonumber
\end{align}
The same approach as the above one for the product over $j$ leads to
the expression promised to be proved. For completeness, let us finish introducing some standard notation:
\begin{align}
&\kappa(\lambda)=2\sum_{(i,j)\in\lambda}(j-i) \\
&\Arrowvert\lambda\Arrowvert^{2}=\sum_{i}\lambda_{i}^{2}\\
&\kappa(\lambda)=\Arrowvert\lambda\Arrowvert^{2}-\Arrowvert\lambda^{t}\Arrowvert^{2}
\end{align}


\begin{thebibliography}{99}
\bibitem{Iqbal:2002we}
A.~Iqbal, ``{All genus topological string amplitudes and 5-brane webs as
  Feynman diagrams},''
\href{http://arXiv.org/abs/hep-th/0207114}{{\tt hep-th/0207114}}.

\bibitem{AKMV}
M.~Aganagic, A.~Klemm, M.~Marino, and C.~Vafa, ``{The topological vertex},''
  {\em Commun. Math. Phys.} {\bf 254} (2005) 425--478,
\href{http://arXiv.org/abs/hep-th/0305132}{{\tt hep-th/0305132}}.

\bibitem{borodin}
A.~Borodin, ``{Periodic Schur process and cylindric partitions},''
  \href{http://arXiv.org/abs/{math/0601019}}{{\tt {math/0601019}}}.

\bibitem{wip}
A.~Iqbal and C.~Kozcaz, ``{U(N) Adjoint Theory, Cylindric Partitions and
  Hilbert Schemes},'' \href{http://arXiv.org/abs/work in progress}{{\tt work in
  progress}}.

\bibitem{carlsson}
E.~Carlsson and A.~Okounkov, ``{Exts and Vertex Operators},''
  \href{http://arXiv.org/abs/{arXiv.org:0801.2565}}{{\tt  {arXiv.org:0801.2565}}}.

\bibitem{Katz:1997eq}
S.~Katz, P.~Mayr, and C.~Vafa, ``{Mirror symmetry and exact solution of 4D N =
  2 gauge theories. I},'' {\em Adv. Theor. Math. Phys.} {\bf 1} (1998) 53--114,
\href{http://arXiv.org/abs/hep-th/9706110}{{\tt hep-th/9706110}}.

\bibitem{Okounkov}
A.~Okounkov and N.~Reshetikhin, ``{Random skew plane partitions and the Pearcey
  process},''
\href{http://arXiv.org/abs/{math/0503508[math.CO]}}{{\tt
  {math/0503508[math.CO]}}}.

\bibitem{Nekrasov:2002qd}
N.~A. Nekrasov, ``{Seiberg-Witten prepotential from instanton counting},'' {\em
  Adv. Theor. Math. Phys.} {\bf 7} (2004) 831--864,
\href{http://arXiv.org/abs/hep-th/0206161}{{\tt hep-th/0206161}}.

\bibitem{IKV}
A.~Iqbal, C.~Kozcaz, and C.~Vafa, ``{The Refined Topological Vertex},'' {\em JHEP} {\bf 10} (2009) 069
\href{http://arXiv.org/abs/hep-th/0701156}{{\tt hep-th/0701156}}.

\bibitem{ORV}
A.~Okounkov, N.~Reshetikhin, and C.~Vafa, ``{Quantum Calabi-Yau and classical
  crystals},''
\href{http://arXiv.org/abs/hep-th/0309208}{{\tt hep-th/0309208}}.

\bibitem{Okuda:2004mb}
T.~Okuda, ``{Derivation of Calabi-Yau crystals from Chern-Simons gauge
  theory},'' {\em JHEP} {\bf 03} (2005) 047,
\href{http://arXiv.org/abs/hep-th/0409270}{{\tt hep-th/0409270}}.

\bibitem{Sulkowski:2006jp}
P.~Sulkowski, ``{Crystal model for the closed topological vertex geometry},''
  {\em JHEP} {\bf 12} (2006) 030,
\href{http://arXiv.org/abs/hep-th/0606055}{{\tt hep-th/0606055}}.

\bibitem{Donagi:1995cf}
R.~Donagi and E.~Witten, ``Supersymmetric Yang-Mills Theory And Integrable Systems,'' {\em Nucl. Phys.}  {\bf B497}, 299 (1996) \href{http://arXiv.org/abs/hep-th/9510101}{{\tt arXiv:hep-th/9510101}}.

\bibitem{Witten:1997sc}
E.~Witten, ``Solutions of four-dimensional field theories via M-theory,'' {\rm Nucl. Phys.}  {\bf B500}, 3 (1997) \href{http://arXiv.org/abs/hep-th/9703166}{{\tt hep-th/9703166}}.

\bibitem{Hollowood:2003cv}
T.~J. Hollowood, A.~Iqbal, and C.~Vafa, ``{Matrix Models, Geometric Engineering
  and Elliptic Genera},''
\href{http://arXiv.org/abs/hep-th/0310272}{{\tt hep-th/0310272}}.

\bibitem{Intriligator:1997pq}
K.~A. Intriligator, D.~R. Morrison, and N.~Seiberg, ``{Five-dimensional
  supersymmetric gauge theories and degenerations of Calabi-Yau spaces},'' {\em
  Nucl. Phys.} {\bf B497} (1997) 56--100,
\href{http://arXiv.org/abs/hep-th/9702198}{{\tt hep-th/9702198}}.

\bibitem{macdonald}
I.~G. Macdonal, ``Symmetric functions and hall polynomials,'' {\em Oxford
  Mathematical Monographs, Oxford Science Publications}
(second edition, 1995).


\bibitem{Nekrasov:2003rj}
N.~Nekrasov and A.~Okounkov, ``{Seiberg-Witten theory and random partitions},''
\href{http://arXiv.org/abs/{hep-th/0306238}}{{\tt {hep-th/0306238}}}.

\bibitem{kerov}
S.~Kerov, ``{Anisotropic Young diagrams and Jack symmetric functions},''. {\tt arXiv:math/9712267}.

\bibitem{GK}
I.~Gessel and C.~Krattenthaler, ``{Cylindric partitions},'' {\em Trans. Amer.
  Math. Soc.} {\bf 349} (1997) no. 2, 429--479.

\bibitem{Haiman}
M.~Haiman, ``{Notes on Macdonald polynomials and the geometry of Hilbert
  schemes},'' {\em {Symmetric Functions 2001: Surveys of Developments and
  Perspectives, Proceedings of the NATO Advanced Study Institute held in
  Cambridge, June 25-July 6, 2001, Sergey Fomin, editor. Kluwer, Dordrecht
  (2002), 1-64,}}.


\bibitem{wip2}
A. Iqbal, work in progress.




\end{thebibliography}
\end{document}